\documentclass[onecolumn,fleqn,usenatbib]{mnras}

\usepackage[T1]{fontenc}

\DeclareRobustCommand{\VAN}[3]{#2}
\let\VANthebibliography\thebibliography
\def\thebibliography{\DeclareRobustCommand{\VAN}[3]{##3}\VANthebibliography}

\usepackage{graphicx}	
\usepackage{amsmath}	
\usepackage{amssymb}	

\title[NFDG AND ROTATIONALLY SUPPORTED GALAXIES]{NEWTONIAN FRACTIONAL-DIMENSION GRAVITY AND ROTATIONALLY SUPPORTED GALAXIES}

\author[G. U. Varieschi]{
Gabriele U. Varieschi,$^{1}$\thanks{E-mail: gvarieschi@lmu.edu}
\\
$^{1}$Department of Physics, Loyola Marymount University, 1 LMU Drive, Los Angeles, CA 90045, USA
}

\date{Accepted XXX. Received YYY; in original form ZZZ}

\pubyear{2020}

\begin{document}
\label{firstpage}
\pagerange{\pageref{firstpage}--\pageref{lastpage}}
\maketitle

\begin{abstract}
We continue our analysis of Newtonian Fractional-Dimension Gravity, an extension of the standard laws of Newtonian gravity to lower dimensional spaces including those with fractional
(i.e., non-integer) dimension. We apply our model to three rotationally supported galaxies: NGC 7814 (Bulge-Dominated Spiral), NGC 6503 (Disk-Dominated Spiral), and NGC 3741 (Gas-Dominated Dwarf).

As was done in the general cases of spherically-symmetric and axially-symmetric structures, which were studied in previous work on the subject, we examine a possible connection between our model and Modified Newtonian Dynamics, a leading alternative
gravity model which explains the observed properties of these galaxies without requiring the Dark Matter hypothesis.

In our model, the MOND acceleration constant
$a_{0} \simeq 1.2 \times 10^{ -10}\mbox{m}\thinspace \mbox{s}^{ -2}$
can be related to a natural scale length
$l_{0}$, namely
$a_{0} \approx GM/l_{0}^{2}$ for a galaxy of mass
$M$. Also, the empirical Radial Acceleration
Relation, connecting the observed radial acceleration
$g_{obs}$ with the baryonic one
$g_{bar}$, can be explained in terms of a variable local dimension $D$. As an example of this methodology, we provide detailed rotation curve fits for the three galaxies mentioned above.
\end{abstract}




\section{Introduction}
\label{sect:intro}
This work continues the analysis started in previous publications (\citet{Varieschi:2020ioh, Varieschi:2020dnd} - paper I and II, respectively, in the following) of \textit{Newtonian Fractional-Dimension Gravity} (NFDG), an extension of the laws of Newtonian gravitation to lower dimensional spaces, including those with non-integer, ``fractional'' dimension (see also \citet{Varieschi:2020}, for a general introduction to NFDG). 

This model is loosely based on the methods of fractional mechanics/electromagnetism (see \citet{bookTarasov,bookZubair,Varieschi:2018}, and references therein) and on the general framework of fractional calculus
(FC) \citep{MR0361633,Herrmann:2011zza}. Several other fractional models of gravity exist in the literature \citep{Muslih2007,Rousan2002,Munkhammar:2010gq,Calcagni:2009kc,Calcagni:2011kn,Calcagni:2011sz,Calcagni:2013yqa,Calcagni:2016azd,Calcagni:2016xtk,Calcagni:2018dhp,Svozil:2017ybx,Giusti:2020rul,Giusti:2020kcv} with most of them directly based on FC, since they use fractional derivatives and fractional operators in their field equations.

Our NFDG as introduced in papers I-II is not a fractional theory, in the sense used by these other models \citep{Giusti:2020rul,Giusti:2020kcv,Calcagni:2016azd}, since the NFDG field equations as outlined in the next section are of integer order; thus local, as opposed to non-local field equations based on fractional differential operators. In paper I, we based NFDG on a generalization of the gravitational Gauss's law, replacing standard space integration over $\mathbb{R}^{3}$ with an appropriate Hausdorff measure over the space, which was related to Weyl's fractional integrals. In this way, Newtonian gravity  was extended to fractal spaces of non-integer dimension, with focus on spherically-symmetric galactic structures, such as homogeneous spheres, Plummer models, etc. In paper II, we extended our analysis to axially-symmetric structures, such as exponential thin/thick disk galaxies, Kuzmin models, etc., and produced a preliminary rotation curve fit for NGC 6503.

Obviously, the goal of NFDG is to describe galactic dynamics without using the controversial DM component, in view of the absence of any direct detection of dark matter in recent years \citep{Bertone:2016nfn,10.1093/ptep/ptaa104}. As shown in papers I-II, NFDG might also be connected with Modified Newtonian Dynamics
(MOND) \citep{Milgrom:1983ca,Milgrom:1983pn,Milgrom:1983zz}, although NFDG is an inherently linear theory due to the integer order of its field equations, while MOND is a fully non-linear theory. In addition, the empirical correlation between the observed gravitational acceleration $g_{obs}$, traced by galactic
rotation curves, and the predicted acceleration $g_{bar}$, based on the observed distribution of baryons (Radial Acceleration Relation - RAR \citet{McGaugh:2016leg,Lelli:2017vgz,Chae:2020omu}) was explained by NFDG in terms of an effective variable dimension $D$, which characterizes each galactic structure.

In Sect. \ref{sect:REVIEW}, we will review the fundamental ideas of NFDG from our first two papers and their connections with MOND/RAR. In Sect. \ref{sect::galactic}, we will apply NFDG to three rotationally supported galaxies, which will be used as reference cases in this work: NGC 7814 (bulge-dominated spiral), NGC 6503 (disk-dominated spiral), and NGC 3741 (gas-dominated dwarf). Also, in Sect. \ref{subsect:discussion} we will discuss how the analysis of these three reference galaxies might be generalized to all rotationally supported galaxies. Finally, in
Sect. \ref{sect::conclusion} we will outline possible future work on the subject and draw our conclusions.

\section{Quick review of NFDG}
\label{sect:REVIEW}
In this section we summarize the main ideas of NFDG and its applications to galactic dynamics. Full details can be found in papers I-II \citep{Varieschi:2020ioh, Varieschi:2020dnd} and references therein. 

Newtonian Fractional-Dimension Gravity was introduced heuristically by extending Gauss's law for gravitation to a lower-dimensional space-time $D +1$, where $D \leq 3$ can be a non-integer space dimension. A scale length $l_{0}$ is needed to ensure dimensional correctness of all expressions for $D \neq 3$; thus, it is convenient to use dimensionless coordinates in all formulas, such as the radial distance $w_{r} \equiv r/l_{0}$ or, in general, the dimensionless coordinates
$\mathbf{w} \equiv \mathbf{x}/l_{0}$
for the field point and
$\mathbf{w}^{ \prime } \equiv \mathbf{x}^{ \prime }/l_{0}$
for the source point. We also introduced a rescaled mass ``density''
$\widetilde{\rho }\left (\mathbf{w}^{ \prime }\right ) =\rho \left (\mathbf{w}\mathbf{}^{ \prime }l_{0}\right )l_{0}^{3} =\rho \left (\mathbf{x}^{ \prime }\right )l_{0}^{3}$, where
$\rho (\mathbf{x}^{ \prime })$
is the standard mass density in
$\mbox{kg}\thinspace \mbox{m}^{ -3}$, with $d\widetilde{m}_{\left (D\right )} =\widetilde{\rho }\left (\mathbf{w}^{ \prime }\right )d^{D}\mathbf{w}^{ \prime }$ representing the infinitesimal mass in a D-dimensional space.\protect\footnote{
	SI units will be used throughout this paper, unless otherwise noted.
}

The NFDG gravitational potential
$\widetilde{\phi }\left (\mathbf{w}\right )$
was introduced as:
\begin{gather}\widetilde{\phi }(\mathbf{w}) = -\frac{2\pi ^{1 -D/2}\Gamma (D/2)G}{\left (D -2\right )l_{0}}{\displaystyle\int _{V_{D}}}\frac{\widetilde{\rho }(\mathbf{w}^{ \prime })}{\left \vert \mathbf{w} -\mathbf{w}^{ \prime }\right .\vert ^{D -2}}d^{D}\mathbf{w}^{ \prime };\ D \neq 2 \label{eq2.1} \\
	\widetilde{\phi }\left (\mathbf{w}\right ) =\frac{2G}{l_{0}}{\displaystyle\int _{V_{2}}}\widetilde{\rho }\left (\mathbf{w}^{ \prime }\right )\ln \left \vert \mathbf{w} -\mathbf{w}^{ \prime }\right .\vert d^{2}\mathbf{w}^{ \prime };\ D =2 \nonumber \end{gather}
with
$\widetilde{\phi }(\mathbf{w})$
and
$\mathbf{g}(\mathbf{w})$
connected by
$\mathbf{g}(\mathbf{w}) = - \nabla _{D}\widetilde{\phi }(\mathbf{w})/l_{0}$, where the D-dimensional gradient
$ \nabla _{D}$
is equivalent to the standard one, but the derivatives are taken with respect to the rescaled coordinates $\mathbf{w}$. It is easy to check that all the expressions above correctly reduce to the standard Newtonian ones for $D =3$.

In our previous papers I-II, we used $\phi \left (\mathbf{w}\right ) \equiv \widetilde{\phi }\left (\mathbf{w}\right )/l_{0}$ to denote the NFDG potentials. Here we prefer to use $\widetilde{\phi }$, so that the first line in Eq. (\ref{eq2.1}) for $D =3$ will yield the standard Newtonian potential. Since $ \nabla _{D}$ is defined in terms of dimensionless coordinates, the physical dimensions for
$\phi $ are the same as
those for the gravitational field
$\mathbf{g}$, i.e., both quantities are measured in $\mbox{m}\thinspace \mbox{s}^{ -2}$. On the contrary,  the physical dimensions for the gravitational potential
$\widetilde{\phi }$ are now the same as those of the standard Newtonian potential (i.e., measured in $\mbox{m}^{2}\thinspace \mbox{s}^{ -2}$).

It should be noted that the fractional dimension $D$ that is used in NFDG is actually the Hausdorff dimension of the matter distribution of the galactic structures being considered. This is evidenced by the integration techniques over D-dimensional metric spaces used in NFDG, which are based on appropriate Hausdorff measures over these spaces (see papers I-II and references therein). Although the potential in Eq. (\ref{eq2.1}) produces an effective modification of gravity for $D\neq3$, NFDG does not imply a change of the tridimensionality of space in galaxies. For example, in our Galaxy, the very tight bounds on the dimensionality of space set by the Cassini probe for the solar system \citep{Will:2014kxa} would not be in contradiction with a local Hausdorff dimension $D\neq3$ for the matter distribution.

The gravitational potential in Eq. (\ref{eq2.1}) was derived for a fixed value of the fractional dimension $D$, but we argued in our previous papers that it could be applicable also to the case of a variable dimension $D\left (\mathbf{w}\right )$, assuming a slow change of the dimension $D$ with the field point coordinates. More rigorous ways to achieve a varying dimension will be discussed in Appendix \ref{sect::appendix}.

The scale length $l_{0}$ was related to the MOND\  acceleration constant
$a_{0}$ (also denoted by $g_{\dag }$ in \citet{McGaugh:2016leg,Lelli:2017vgz,Chae:2020omu}):

\begin{equation}a_{0} \equiv g_{\dag } =1.20 \pm 0.02\ \text{(random)} \pm 0.24\ \text{(syst)} \times 10^{ -10}\ \mbox{}\ \mbox{m}\thinspace \mbox{s}^{ -2} , \label{eq2.2}
\end{equation}
which represents the acceleration scale below which MOND corrections are needed.

MOND \citep{Milgrom:1983ca,Milgrom:1983pn,Milgrom:1983zz} proposed modifications of Newtonian dynamics in terms of modified inertia (MI), or modified gravity (MG)
\citep{Bekenstein:1984tv}. Respectively:
$m\mu (a/a_{0})\mathbf{a} =\mathbf{F}$ as MI, since the mass $m$ is replaced by $m\mu \left (a/a_{0}\right )$, and $\mu (g/a_{0})\mathbf{g} =\mathbf{g}_{N}$ as MG, since the observed gravitational field $\mathbf{g}$ can differ from the Newtonian one $\mathbf{g}_{N}$. While there is now limited evidence \citep{Petersen:2020vks,Milgrom:2012rk} that MG might be favored over MI, the two formulations are practically equivalent, although conceptually
different:\ the former modifies Newton's laws of motion, while
the latter modifies Newton's law of universal gravitation. Following Eq. (\ref{eq2.1}) above, NFDG can be also considered an effective modification of gravity and not of inertia, since we assume that test objects, such as stars in galaxies, will still move in a
(classical)
$3 +1$
space-time and obey standard laws of dynamics. 

The \textit{interpolation function, }$\mu (x) \equiv \mu (a/a_{0})\text{}$
for the MI\  case or
$\mu \text{}(x) \equiv \mu (g/a_{0})$ for the MG case, was introduced by MOND as: $\mu \left (x\right ) \rightarrow 1\text{  for }x \gg 1\text{  (Newtonian regime)}$ and $\mu \left (x\right ) \rightarrow x$ $\text{for }x \ll 1\text{  (Deep-MOND limit)}$, thus defining only the two asymptotic MOND\ behaviors. The bridge between these two limiting behaviors is usually represented by an explicit interpolation function $\mu \left (x\right )$ (or its inverse function $\nu \left (y\right )$, see papers I-II for details), with the particular choice
$\nu (y) =\left [1 -\exp \left ( -y^{1/2}\right )\right ]^{ -1}$ as the favorite interpolation function in the literature \citep{McGaugh:2016leg,Lelli:2017vgz,Chae:2020omu}. This function is equivalent to the \textit{Radial Acceleration Relation} - RAR:

\begin{equation}g_{obs} =\frac{g_{bar}}{1 -e^{ -\sqrt{g_{bar}/g_{\dag }}}} , \label{eq2.3}
\end{equation}
where
$g_{\dag }$
corresponds numerically to the MOND
acceleration scale
$a_{0}$, reported in Eq. (\ref{eq2.2})
above.

The RAR relates the
radial acceleration
$g_{obs}$
traced by rotation curves with the radial acceleration
$g_{bar}$
predicted by the observed distribution of baryonic matter in galaxies \citep{McGaugh:2016leg,Lelli:2017vgz,Chae:2020omu}  and was originally obtained by analyzing data from 175 galaxies in the Spitzer Photometry and Accurate Rotation Curves
(SPARC) database \citep{Lelli:2016zqa}. It was confirmed in later work \citep{Lelli:2017vgz,Li:2018tdo} by adding
early-type-galaxies (elliptical and lenticular) and dwarf spheroidal galaxies to the same database. Additionally, a detection of the external field effect, which is typical of Milgromian dynamics (MOND), has been recently reported \cite{Chae:2020omu}.

In papers I-II, we proposed a possible connection between the scale length $l_{0}$ and the MOND\ acceleration $a_{0}$ as:
\begin{equation}a_{0} \approx \frac{GM}{l_{0}^{2}} , \label{eq2.4}
\end{equation}where $M$ is the total mass (or a reference mass) of the system being studied. It was shown \ that the main consequences of the MOND\ theory could be recovered in NFDG by considering the Deep-MOND limit equivalent to reducing the space dimension to $D \approx 2$. In particular, the asymptotic or
flat rotation velocity
$V_{f} \approx \sqrt[{4}]{GMa_{0}}$ exhibited by galactic rotation curves, the ``baryonic'' Tully-Fisher relation-BTFR:
$M_{bar} \sim V_{f}^{4}$, and other fundamental MOND predictions were recovered in NFDG for the case $D \approx 2$  \citep{Varieschi:2020ioh}.

In paper I, we mainly studied spherically-symmetric mass distributions $\widetilde{\rho }\left (w_{r}^{ \prime }\right )$ and proved that the gravitational field
$\mathbf{g}(w_{r})$, in a fractal space of dimension
$D(w_{r}^{})$ depending on the radial distance $w_{r} =r/l_{0}$, can be computed
as:

\begin{equation}\mathbf{g}_{obs}(w_{r}) = -\frac{4\pi G}{l_{0}^{2}w_{r}^{D\left (w_{r}\right ) -1}}{\displaystyle\int _{0}^{w_{r}}}\tilde{\rho }\left (w_{r}^{ \prime }\right )w_{r}^{ \prime ^{D\left (w_{r}\right ) -1}}dw_{r}^{ \prime }\overset{}{\widehat{\mathbf{w}}_{r} ,} \label{eq2.5}
\end{equation}
for
$1 \leq D \leq 3$. In the last equation, the gravitational field is denoted as the
``observed'' one,
$\mathbf{g}_{obs}$, while the ``baryonic''
$\mathbf{g}_{bar}$ is considered to be the one for  fixed dimension
$D =3$: 

\begin{equation}\mathbf{g}_{bar}(w_{r}) = -\frac{4\pi G}{l_{0}^{2}w_{r}^{2}}{\displaystyle\int _{0}^{w_{r}}}\tilde{\rho }\left (w_{r}^{ \prime }\right )w_{r}^{ \prime ^{2}}dw_{r}^{ \prime }\overset{}{\widehat{\mathbf{w}}_{r} .} \label{eq2.6}
\end{equation}

Therefore, the connection with MOND and the RAR was simply obtained in NFDG by considering the ratio between the previous two equations:

\begin{equation}\genfrac{(}{)}{}{}{g_{obs}}{g_{bar}}_{NFDG}(w_{r}) =w_{r}^{3 -D\left (w_{r}\right )}\frac{{\displaystyle\int _{0}^{w_{r}}}\widetilde{\rho }\left (w_{r}^{ \prime }\right )w_{r}^{ \prime ^{D\left (w_{r}\right ) -1}}dw_{r}^{ \prime }}{{\displaystyle\int _{0}^{w_{r}}}\widetilde{\rho }\left (w_{r}^{ \prime })\right .w_{r}^{ \prime ^{2}}dw_{r}^{ \prime }} , \label{eq2.7}
\end{equation}and by comparing it with the similar ratio coming from the RAR in Eq. (\ref{eq2.3}):
$\genfrac{(}{)}{}{}{g_{obs}}{g_{bar}}_{MOND}\left (w_{r}\right ) =\frac{1}{1 -e^{ -\sqrt{g_{bar}\left (w_{r}\right )/g_{\dag }}}}$.

This procedure was indeed successful for several different forms of spherically-symmetric mass distributions analyzed in paper I. In each case, we computed numerically the dimension functions
$D(w_{r}^{})$ by comparing directly the ratios           $\genfrac{(}{)}{}{}{g_{obs}}{g_{bar}}_{NFDG}$ and           $\genfrac{(}{)}{}{}{g_{obs}}{g_{bar}}_{MOND}$ above. The variable dimension $D\left (w_{r}\right )$ assumed values
$D \approx 3$
in regions where Newtonian gravity appeared to hold, while it decreased continuously
toward $D \approx 2$ in Deep-MOND limit regions.

In paper II, we applied NFDG to the case of axially-symmetric mass distributions. Instead of computing directly the gravitational field, as was done for the spherically-symmetric case, we first calculated the gravitational potential using Eq. (\ref{eq2.1}) and then used the fractional gradient
$ \nabla _{D}$ to obtain the field. For the general NFDG potential in the first line of Eq. (\ref{eq2.1}), $\widetilde{\phi } \sim 1/\left \vert \mathbf{w} -\mathbf{w}^{ \prime }\right \vert ^{D -2}$ (sometimes referred to as the \textit{Euler kernel}), we used the following expansion \citep{2012JPhA...45n5206C} in (rescaled) spherical coordinates:

\begin{equation}\frac{1}{\left \vert \mathbf{w} -\mathbf{w}^{ \prime }\right \vert ^{D -2}} =\sum \limits _{l =0}^{\infty }\frac{w_{r<}^{l}}{w_{r>}^{l +D -2}}C_{l}^{\left (\frac{D}{2} -1\right )}\left (\cos \gamma \right ) , \label{eq2.8}
\end{equation}
where $w_{r<}$ ($w_{r>}$) is the smaller (larger) of $w_{r}$ and $w_{r}^{ \prime }$, $\gamma $ is the angle between the unit vectors $\widehat{\mathbf{w}}$ and $\widehat{\mathbf{w}}^{ \prime }$, and $C_{l}^{\left (\lambda \right )}\left (x\right )$ denotes Gegenbauer polynomials (see paper I or \citet{NIST} for general properties of these special functions).

This expansion was easily adapted to the case of cylindrical coordinates ($w_{R}$, $\varphi $, $w_{z}$). In the case of thin disks, in the $w_{z} =w_{z}^{ \prime } =0$ plane and in the $\varphi  =0$ direction, the angle $\gamma $ is replaced by $\varphi ^{ \prime }$ and the radial spherical coordinate $w_{r} \equiv r/l_{0}$ with the cylindrical $w_{R} \equiv R/l_{0}$:
\begin{equation}\frac{1}{\left \vert \mathbf{w} -\mathbf{}\mathbf{w}^{ \prime }\right \vert ^{D -2}} =\sum \limits _{l =0}^{\infty }\frac{w_{R<}^{l}}{w_{R>}^{l +D -2}}C_{l}^{\left (\frac{D}{2} -1\right )}\left (\cos \varphi ^{ \prime }\right ) , \label{eq2.9}
\end{equation}
while, for thick disks with $w_{z}^{ \prime } \neq 0$ (but still in the $w_{z} =0$ plane and $\varphi  =0$ direction), we used the following coordinate transformations:

\begin{gather}w_{r} =w_{R} \label{eq2.10} \\
	w_{r}^{ \prime } =\sqrt{w_{R}^{ \prime 2} +w_{z}^{ \prime {2}}} \nonumber  \\
	\cos \gamma  =\frac{w_{R}^{ \prime }\cos \varphi ^{ \prime }}{\sqrt{w_{R}^{ \prime {2}} +w_{z}^{ \prime {2}}}} , \nonumber \end{gather}and modified the original expansion in Eq.  (\ref{eq2.8}) accordingly.

The general NFDG potential in the first line of Eq. (\ref{eq2.1}) was then computed using the techniques for multi-variable integration over a fractal metric space $W \subset \mathbb{R}^{3}$ (see again papers I-II), combined with the expansions of the Euler kernel outlined above. For thin/thick disk structures, the rescaled mass distributions were taken, respectively, as:
\begin{gather}\widetilde{\rho }\left (w_{R}^{ \prime } ,w_{z}^{ \prime }\right ) =\widetilde{\Sigma }\left (w_{R}^{ \prime }\right )\delta \left (w_{z}^{ \prime }\right ) \label{eq2.11} \\
	\widetilde{\rho }\left (w_{R}^{ \prime } ,w_{z}^{ \prime }\right ) =\widetilde{\Sigma }\left (w_{R}^{ \prime }\right )\widetilde{\zeta }\left (w_{z}^{ \prime }\right ) \nonumber \end{gather}where the surface mass distribution $\widetilde{\Sigma }\left (w_{R}^{ \prime }\right )$ can be related to an exponential model, a Kuzmin model, or simply obtained by interpolating SPARC surface luminosity data and transforming them into surface mass distributions, using appropriate mass-to-light ratios.

For thin disks, the vertical density is simply described by the delta function $\delta \left (w_{z}^{ \prime }\right )$, while for thick disks we typically use an exponential function  $\widetilde{\zeta }\left (w_{z}^{ \prime }\right ) =\frac{1}{2H_{z}}e^{ -w_{z}^{ \prime }/H_{z}}$, where the rescaled parameter $H_{z} =h_{z}/l_{0}$ is connected with the original vertical scale height $h_{z}$. We also adopted the standard relation \citep{Lelli:2016zqa,2010ApJ...716..234B}, $\left (h_{z}/\ensuremath{\operatorname*{kpc}}\right ) =0.196\left (R_{d}/\ensuremath{\operatorname*{kpc}}\right )^{0.633}$, between the vertical scale height $h_{z}$ and the radial scale length $R_{d}$ (available from SPARC data), properly rescaled by using our dimensionless variables.

For exponential and Kuzmin thin disks, the NFDG potential can be obtained analytically and full results were reported in paper II, while similar analytic results for spherical models were reported in both papers I and II. In this work, we will analyze instead galaxies whose mass distributions are obtained directly from the SPARC luminosity data. In general, SPARC data include three types of luminosity distributions: a spherically-symmetric \textit{bulge}, a stellar \textit{disk} component, and a \textit{gas} disk component (in the following: bulge, disk, and gas components, for short).

For the disk and gas components, the SPARC surface luminosities $\Sigma _{disk}^{(L)}\left (R\right )$ and $\Sigma _{gas}^{(L)}\left (R\right )$, can be turned into (rescaled) surface mass distributions, $\widetilde{\Sigma }_{disk}\left (w_{R}^{ \prime }\right )$ and $\widetilde{\Sigma }_{gas}\left (w_{R}^{ \prime }\right )$, by using appropriate mass-to-light ratios \citep{Lelli:2020pri}: $\Upsilon _{disk} \simeq 0.50\ M_{ \odot }/L_{ \odot }$, $\Upsilon _{gas} \simeq 1.33\ M_{ \odot }/L_{ \odot }$ (this value for $\Upsilon _{gas}$ includes also the helium gas contribution). We then sum these distributions together, $\widetilde{\Sigma }\left (w_{R}^{ \prime }\right ) =\widetilde{\Sigma }_{gas}\left (w_{R}^{ \prime }\right ) +\widetilde{\Sigma }_{disk}\left (w_{R}^{ \prime }\right )$, and combine this total surface distribution $\widetilde{\Sigma }\left (w_{R}^{ \prime }\right )$ with the vertical exponential function $\widetilde{\zeta }\left (w_{z}^{ \prime }\right )$ described above, into the second line of Eq. (\ref{eq2.11}).

For the bulge component, SPARC data are also available in terms of a surface luminosity $\Sigma _{bulge}^{(L)}\left (R\right )$ which can be turned into a (rescaled) surface mass distribution $\widetilde{\Sigma }_{bulge}\left (w_{R}^{ \prime }\right )$ by using $\Upsilon _{bulge} \simeq 0.70\ M_{ \odot }/L_{ \odot }$ \citep{Lelli:2020pri} and then converted into a spherically-symmetric mass distribution $\tilde{\rho }\left (w_{r}^{ \prime }\right )^{}$ by applying Eq. (1.79) in \citet{2008gady.book.....B}. In this way, for each galaxy studied in this paper, we will have bulge, disk, and gas mass distributions based directly on the experimental SPARC data, instead of using pre-determined models, such as exponential, Kuzmin for disk galaxies, and Plummer or other models for spherical components.

For the combined disk and gas components, the NFDG potential from Eq. (\ref{eq2.1}) can be obtained by applying the expansion in Eqs. (\ref{eq2.8})-(\ref{eq2.10}) and by performing a triple integration over $w_{R}^{ \prime }$, $\varphi ^{ \prime }$, and $w_{z}^{ \prime }$. As discussed in paper II, we also need to connect the overall space dimension $D$ with the respective fractional dimensions - $\alpha _{R}$, $\alpha _{\varphi }$, $\alpha _{z}$ - of each sub-space, so that $D =\alpha _{R} +\alpha _{\varphi } +\alpha _{z}$. As in paper II, we assume $\alpha _{z} =1$, i.e., no fractional dimension is needed in the $z^{ \prime }$ direction, as opposed to $\alpha _{R} =\alpha _{\varphi } =\alpha  =\frac{D -1}{2}$, so that the radial and angular coordinates will share the same (variable) fractional dimension. Therefore, we will have: $D =2\alpha  +1 \leq 3$, and the results will depend only on the overall dimension $D$ of the space.

With these assumptions, and using the techniques outlined in papers I-II for multi-variable integration over a fractal metric space, we then obtain the total thick-disk potential in the $w_{z} =0$ plane as:

\begin{gather}\widetilde{\phi }\left (w_{R}\right ) = -\frac{4\sqrt{\pi }\ \Gamma \left (D/2\right )G}{\left (D -2\right )\left [\Gamma \genfrac{(}{)}{}{}{D -1}{4}\right ]^{2}l_{0}}\sum \limits _{l =0}^{\infty }{\displaystyle\int _{0}^{\infty }}\widetilde{\Sigma }\left (w_{R}^{ \prime }\right )\frac{w_{R <}^{l}}{w_{R >}^{l +D -2}}w_{R}^{ \prime D -2}dw_{R}^{ \prime }{\displaystyle\int _{0}^{\infty }}\widetilde{\zeta }\left (w_{z}^{ \prime }\right )c_{l ,D}\left (w_{R}^{ \prime } ,w_{z}^{ \prime }\right )dw_{z}^{ \prime } \label{eq2.12} \\
	= -\frac{4\sqrt{\pi }\ \Gamma \left (D/2\right )G}{\left (D -2\right )\left [\Gamma \genfrac{(}{)}{}{}{D -1}{4}\right ]^{2}l_{0}}\sum \limits _{l =0}^{\infty }\Bigg\{\left [{\displaystyle\int _{0}^{w_{R}}}\widetilde{\Sigma }\left (w_{R}^{ \prime }\right )\frac{\left (\sqrt{w_{R}^{ \prime 2} +w_{z}^{ \prime 2}}\right )^{l}}{w_{R}^{l +D -2}}w_{R}^{ \prime D -2}dw_{R}^{ \prime }{\displaystyle\int _{0}^{\infty }}\widetilde{\zeta }\left (w_{z}^{ \prime }\right )c_{l ,D}\left (w_{R}^{ \prime } ,w_{z}^{ \prime }\right )dw_{z}^{ \prime }\right ]  \nonumber  \\
	+\left [{\displaystyle\int _{w_{R}}^{\infty }}\widetilde{\Sigma }\left (w_{R}^{ \prime }\right )\frac{w_{R}^{l}}{\left (\sqrt{w_{R}^{ \prime 2} +w_{z}^{ \prime 2}}\right )^{l +D -2}}w_{R}^{ \prime D -2}dw_{R}^{ \prime }{\displaystyle\int _{0}^{\infty }}\widetilde{\zeta }\left (w_{z}^{ \prime }\right )c_{l ,D}\left (w_{R}^{ \prime } ,w_{z}^{ \prime }\right )dw_{z}^{ \prime }\right ]\Bigg\} \nonumber \end{gather}

In the previous equation the integrals in $w_{R}^{ \prime }$ and $w_{z}^{ \prime }$ will need to be computed numerically,\protect\footnote{
	All the numerical computations (and some of the analytical ones) in this work, were performed with Mathematica, Version 12.1.1, Wolfram Research Inc.} while we have denoted with $c_{l ,D}\left (w_{R}^{ \prime } ,w_{z}^{ \prime }\right )$ the results of the angular integrations for $D >1$, i.e.:

\begin{gather}c_{l ,D}\left (w_{R}^{ \prime } ,w_{z}^{ \prime }\right ) ={\displaystyle\int _{0}^{2\pi }}\left \vert \sin \varphi ^{ \prime }\right \vert ^{\frac{D -3}{2}}\left \vert \cos \varphi ^{ \prime }\right \vert ^{\frac{D -3}{2}}C_{l}^{\left (\frac{D}{2} -1\right )}\genfrac{(}{)}{}{}{w_{R}^{ \prime }\cos \varphi ^{ \prime }}{\sqrt{w_{R}^{ \prime 2} +w_{z}^{ \prime 2}}}d\varphi ^{ \prime } \label{eq2.13} \\
	c_{0 ,D} = -\frac{2^{\frac{5 -D}{2}}\pi ^{3/2}\sec \genfrac{[}{]}{}{}{\pi \left (1 +D\right )}{4}}{\Gamma \genfrac{(}{)}{}{}{5 -D}{4}\Gamma \genfrac{(}{)}{}{}{1 +D}{4}} ;c_{2 ,D} =\frac{2^{\frac{1 -D}{2}}\pi ^{1/2}(D -2)\left [\left (D -2\right )w_{R}^{ \prime 2} -2w_{z}^{ \prime 2}\right ]\Gamma \genfrac{(}{)}{}{}{D -1}{4}}{\left (w_{R}^{ \prime 2} +w_{z}^{ \prime 2}\right )\Gamma \genfrac{(}{)}{}{}{1 +D}{4}} ; . . . \nonumber  \\
	c_{1 ,D} =c_{3 ,D} = . . . =0 \nonumber \end{gather}where these functions are identically zero for odd values of $l$, while they can be computed analytically for all even values of $l$.

The observed radial acceleration $\mathbf{g}_{obs}$, in the $w_{z} =0$ plane, can be obtained directly from Eqs. (\ref{eq2.12})-(\ref{eq2.13}) by differentiation, i.e., $\mathbf{g}_{obs}\left (w_{R}\right ) = -\frac{d\phi }{dw_{R}}\widehat{\mathbf{w}}_{R}$. This radial acceleration $\mathbf{g}_{obs}$ can be compared with the standard baryonic $\mathbf{g}_{bar}$, obtained with the same procedure but with a fixed $D =3$ value, or directly using the total baryonic rotational velocity, $V_{bar}\left (R\right )$, available from the SPARC data for each galaxy and with $g_{bar}\left (R\right ) =V_{bar}^{2}\left (R\right )/R$. In Sect. \ref{sect::galactic}, we will use this second option to compute $\mathbf{g}_{bar}$, since it allows for a more direct comparison with the Newtonian behavior.

Finally, in order to include also the spherical bulge in our NFDG\ analysis we recall from paper II that the NFDG potential expansion in Eq. (\ref{eq2.8}) is better suited to spherical coordinates, rather than cylindrical. Aligning the field vector $\mathbf{w}$ in the direction of the $w_{z}^{ \prime }$ axis and using (rescaled) spherical coordinates ($w_{r}^{ \prime }$, $\theta ^{ \prime }$, $\varphi ^{ \prime }$) for the source vector $\mathbf{}\mathbf{w}^{ \prime }$, we can use directly the expansion (\ref{eq2.8}) with the angle $\gamma $ replaced by $\theta ^{ \prime }$.

To compute the gravitational potential $\widetilde{\phi }\left (w_{r}\right )$ for a spherically-symmetric mass distribution $\widetilde{\rho }\left (w_{r}^{ \prime }\right )$, we can still use our main Eq. (\ref{eq2.1}), but the volume integration must be performed in spherical coordinates. The full details can be found in our paper II; here we present again the final results obtained by assuming that the fractional dimension will apply to all three coordinates equally, i.e., $\alpha _{r} =\alpha _{\theta } =\alpha _{\varphi } =D/3$, so that the appropriate angular integrals in this case are:
\begin{gather}c_{l ,D} =\int _{0}^{\pi }\left \vert \sin \theta ^{ \prime }\right \vert ^{\frac{2D}{3} -1}\left \vert \cos \theta ^{ \prime }\right \vert ^{\frac{D}{3} -1}C_{l}^{\left (\frac{D}{2} -1\right )}\left (\cos \theta ^{ \prime }\right )d\theta ^{ \prime } \label{eq2.14} \\
	c_{0 ,D} =\frac{\pi \csc \genfrac{(}{)}{}{}{\pi D}{6}\Gamma \genfrac{(}{)}{}{}{D}{3}}{\Gamma \left (1 -\frac{D}{6}\right )\Gamma \genfrac{(}{)}{}{}{D}{2}} ;c_{2 ,D} =\frac{\pi \left (\frac{D}{3} -1\right )\csc \genfrac{(}{)}{}{}{\pi D}{6}\Gamma \left (\frac{D}{3}\right )}{\Gamma \left (1 -\frac{D}{6}\right )\Gamma \left (\frac{D}{2} -1\right )} ; . . . \nonumber  \\
	\int _{0}^{2\pi }\left \vert \sin \varphi ^{ \prime }\right \vert ^{\frac{D}{3} -1}\left \vert \cos \varphi ^{ \prime }\right \vert ^{\frac{D}{3} -1}d\varphi ^{ \prime } =\frac{2\pi \csc \genfrac{(}{)}{}{}{\pi D}{6}\Gamma \genfrac{(}{)}{}{}{D}{6}}{\Gamma \left (1 -\frac{D}{6}\right )\Gamma \genfrac{(}{)}{}{}{D}{3}} . \nonumber \end{gather}

The general potential for spherically-symmetric mass distributions can be written as:

\begin{gather}\widetilde{\phi }\left (w_{r}\right ) = -\frac{2\pi \Gamma \left (\frac{D}{2} -1\right )G}{\Gamma \left (\frac{D}{3}\right )\Gamma \genfrac{(}{)}{}{}{D}{6}l_{0}}\sum \limits _{l =0 ,2 ,4 , . . .}^{\infty }c_{l ,D}{\displaystyle\int _{0}^{\infty }}\widetilde{\rho }\left (w_{r}^{ \prime }\right )\frac{w_{r <}^{l}}{w_{r >}^{l +D -2}}w_{r}^{ \prime D -1}dw_{r}^{ \prime } \label{eq2.15} \\
	= -\frac{2\pi \Gamma \left (\frac{D}{2} -1\right )G}{\Gamma \left (\frac{D}{3}\right )\Gamma \genfrac{(}{)}{}{}{D}{6}l_{0}}\sum \limits _{l =0 ,2 ,4 , . . .}^{\infty }c_{l ,D}\left ({\displaystyle\int _{0}^{w_{r}}}\widetilde{\rho }\left (w_{r}^{ \prime }\right )\frac{w_{r}^{ \prime l}}{w_{r}^{l +D -2}}w_{r}^{ \prime D -1}dw_{r}^{ \prime } +{\displaystyle\int _{w_{r}}^{\infty }}\widetilde{\rho }\left (w_{r}^{ \prime }\right )\frac{w_{r}^{l}}{w^{ \prime l +D -2}_{r}}w_{r}^{ \prime D -1}dw_{r}^{ \prime }\right ) \nonumber \end{gather}and the observed radial acceleration $\mathbf{g}_{obs}\left (w_{r}\right )$ can be obtained again by differentiation of the last equation. The standard radial acceleration $\mathbf{g}_{bar}$ can be simply obtained from Eqs. (\ref{eq2.14})-(\ref{eq2.15}) with $D =3$, or derived directly from SPARC data as already remarked above for the disk case.

For galaxies such as NGC 7814 (see Sect. \ref{subsect:NGC7814}) whose mass distribution consists of a spherical bulge plus cylindrical disk and gas components, we will simply add together the NFDG potentials from Eqs. (\ref{eq2.12}) and (\ref{eq2.15}), and then obtain $\mathbf{g}_{obs}\left (w_{r}\right )$ by differentiation. On the contrary, NGC 6503 (see Sect. \ref{subsect:NGC6503}) and NGC 3741 (see Sect. \ref{subsect:NGC3741}) do not possess a spherical bulge component, so they will be modeled using just the NFDG potential in Eq. (\ref{eq2.12}).

As a final consideration, we note that all our formulas for the gravitational potential $\widetilde{\phi }$ (and, thus, also for the gravitational field $\mathbf{g}_{obs}$) require summations over all non-zero terms for $l =0 ,2 ,4 , . . .$ (terms for odd values of $l$ being identically zero). As in paper II, we found that these series of functions converge rather quickly over most of the range for $w_{R} >0$ (or $w_{r} >0$), with the exception of the thick-disk formulas for very low values of $w_{R}$.

All the results that will be presented in the following sections were computed by summing typically the first six non-zero terms (for $l =0 ,2 ,4 ,6 ,8 ,10$) of our NFDG expansions. For each case, we also checked numerically that our NFDG\ expansions reduce to the standard $\mathbf{g}_{bar}$ values from SPARC data, again by summing only the first few non-zero terms of our ($D =3$) \ expansions. Therefore, we are confident that our NFDG\ formulas can be used to describe accurately the physical reality of galactic structures in spaces of dimension $D \leq 3$.

\section{Galactic data fitting}
\label{sect::galactic} 

In the
following sub-sections, we will apply NFDG to three notable examples of rotationally supported galaxies from the SPARC database: NGC 7814, NGC 6503, and NGC 3741. The choice of these galaxies is simply due to the fact that they were used as the main examples of the RAR in the seminal paper by McGaugh, Lelli, and Schombert \citep{McGaugh:2016leg}. The detailed luminosity data for the gas, disk, and bulge components of these three galaxies were obtained from the SPARC database administrator \citep{Lelli:2020pri} and from the publicly available data \citep{Lelli:2016zqa}.

\subsection{NGC 7814}
\label{subsect:NGC7814}

We will start with the case of NGC 7814, a spiral galaxy approximately 40 million light-years away in the constellation Pegasus. This galaxy, also known as the ``Little Sombrero'' for its similarities with M104 (the ``Sombrero Galaxy''), has a bright central bulge and a bright gas halo extending outward in space, as seen edge-on from Earth in images taken by the NASA/ESA Hubble Space Telescope \citep{NGC7814:2020}.

This is a bulge-dominated galaxy, in which all three components (bulge, disk, and gas) are present, with the spherically-symmetric bulge being the most prominent. Following the general discussion of NFDG methods outlined in Sect. \ref{sect:REVIEW}, we converted the SPARC surface luminosity data for the three components into equivalent mass distributions $\widetilde{\Sigma }\left (w_{R}^{ \prime }\right )$ (disk plus gas) and $\widetilde{\rho }\left (w_{r}^{ \prime }\right )$ (bulge). We then entered these functions into our general NFDG potentials in Eqs. (\ref{eq2.12})-(\ref{eq2.15}), from which we then obtained $\mathbf{}g_{obs}\left (w_{R}\right )$ by differentiation.

Following our NFDG assumptions, we can define
$(g_{obs}/g_{bar})_{NFDG}$
as the ratio of $\mathbf{}g_{obs}$ defined above and the Newtonian $g_{bar}$ from the SPARC data (after converting the total baryonic rotational velocity $V_{bar}\left (R\right )$ into $g_{bar}\left (R\right ) =V_{bar}^{2}\left (R\right )/R$):

\begin{equation}\genfrac{(}{)}{}{}{g_{obs}}{g_{bar}}_{NFDG}(w_{R}) =\frac{g_{obs}\left (w_{R} ,D\left (w_{R}\right )\right )}{g_{bar}\left (w_{R}\right )}^{\left .\right .} . \label{eq3.1}
\end{equation}

The observed gravitational field in the last equation is denoted as $g_{obs}\left (w_{R} ,D\left (w_{R}\right )\right )$ to indicate that the dimension $D$ is now considered a function of the field point radial coordinate, i.e., $D =D(w_{R})$, but not a function of the angular coordinates due to the symmetry of the galactic structure (spherical symmetry of the bulge and cylindrical symmetry of disk/gas components).\protect\footnote{
	Again, more rigorous ways to achieve a varying dimension will be discussed in Appendix \ref{sect::appendix}.

} 

As was done in paper II, the NFDG\ ratio
$\genfrac{(}{)}{}{}{g_{obs}}{g_{bar}}_{NFDG}\left (w_{R}\right )$ above, can be compared with the MOND (RAR) ratio
$\genfrac{(}{)}{}{}{g_{obs}}{g_{bar}}_{MOND}\left (w_{R}\right ) =\frac{1}{1 -e^{ -\sqrt{g_{bar}\left (w_{R}\right )/g_{\dag }}}}$, or directly with the ratio of the experimental\ SPARC data for NGC 7814, $\genfrac{(}{)}{}{}{g_{obs}}{g_{bar}}_{SPARC}\left (w_{R}\right )$, as obtained from the rotational velocity data \citep{Lelli:2016zqa,Lelli:2020pri} transformed into radial accelerations. In the former case, we are able to determine $D(w_{R})$
based on the general MOND (RAR) model, while in the latter case the variable dimension $D(w_{R})$
is based directly on the SPARC - NGC 7814 experimental data, thus allowing for a more direct validation of NFDG methods.


\begin{figure}
	\includegraphics[width=\columnwidth]{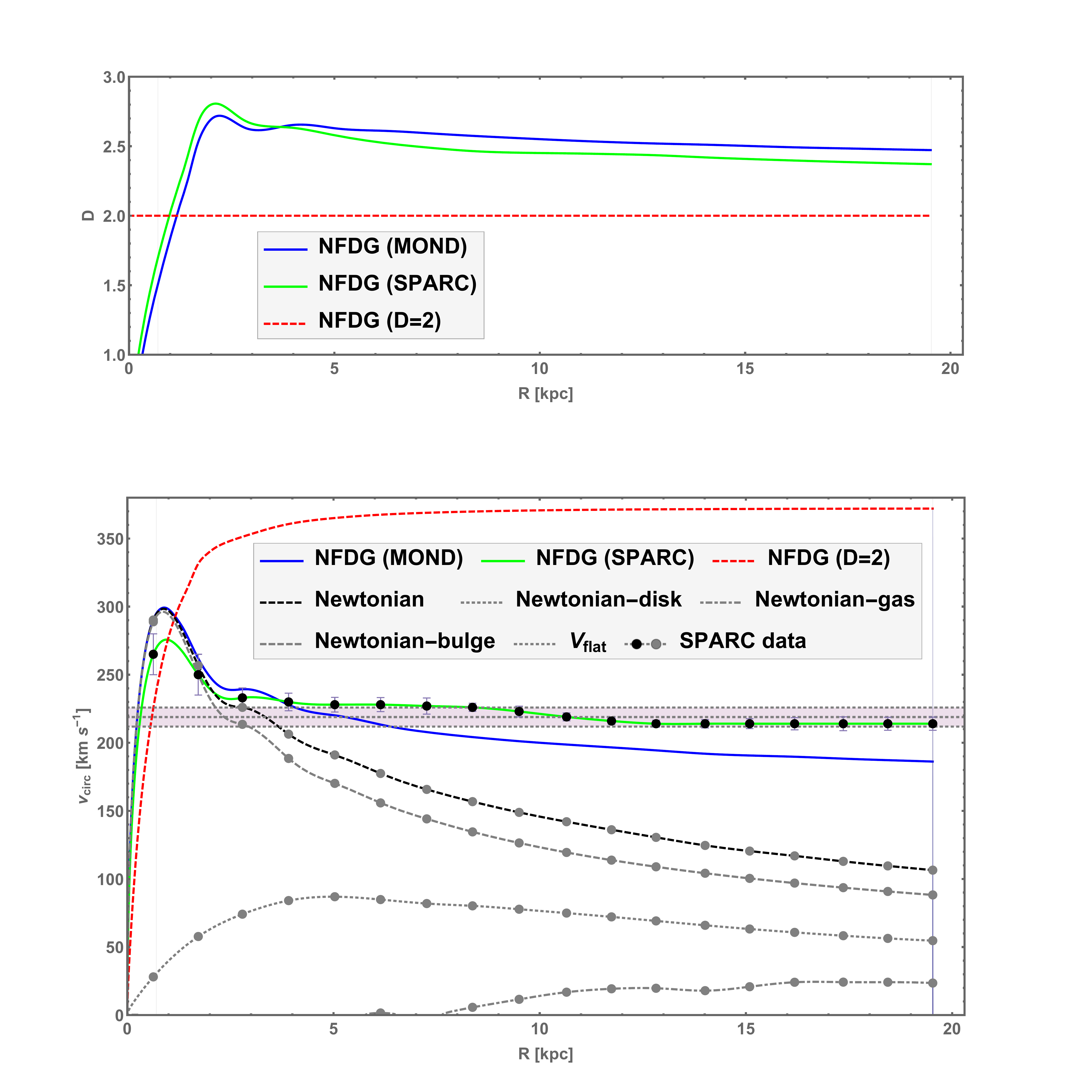}
	\caption{NFDG results for NGC 7814.
		Top panel: NFDG variable dimension $D\left (R\right )$ for MOND (RAR) interpolation function, or based directly on SPARC data. Bottom panel: NFDG rotation curves (circular velocity vs. radial distance) compared to the original SPARC data (black circles). Also shown: Newtonian rotation curves (total and different components - gray lines) and corresponding SPARC data (gray circles).}
	\label{figure:NGC7841_1}
\end{figure}

Figure \ref{figure:NGC7841_1} shows the main NFDG results for NGC 7814. While in papers I-II most of the figures were plotted in terms of rescaled distances $w_{r}$ or $w_{R}$, in this work we prefer to use the astrophysical radial distance $R$ in \textrm{kiloparsec}, while rotational circular velocities $v_{circ}$ are measured in $\mbox{km}\ \mbox{s}^{ -1}$.

The top panel illustrates the variable dimension
$D\left (R\right )$
obtained by equating the NFDG ratio $\left (g_{obs}/g_{bar}\right )_{NFDG}$  in Eq. (\ref{eq3.1}) with the similar MOND (RAR) and SPARC ratios, respectively. The resulting curves (blue-solid for MOND, green-solid for SPARC) show $D \approx 2.7 -2.8$ at small radial distances and then a decrease toward $D \approx 2.4 -2.5$ at larger distances. In this panel, we also show the NFDG $D =2$ reference value (red-dashed), since we argued in Sect. \ref{sect:REVIEW} that this dimension value represents the deep-MOND limit.

At the smallest radial distances, our NFDG results are extrapolated and, therefore, not fully reliable. We already mentioned in paper II that our numerical computation method, based on the expansion in Eqs. (\ref{eq2.8})-(\ref{eq2.10}), has some convergence issues at low radial distances. The minimum distance below which our results were extrapolated is indicated by vertical gray-thin lines in both panels in Fig. \ref{figure:NGC7841_1} at $R_{\min } \simeq 0.70\ \ kpc$ (the same considerations will apply to the similar figures for the other galaxies in Sects. \ref{subsect:NGC6503}-\ref{subsect:NGC3741}). Other vertical gray-thin lines at $R_{\max } \simeq 19.53\ kpc$ limit our plots at large distances, corresponding to the radial distance of the last SPARC data-point for this galaxy.

The low-distance results are also affected by the uncertainties in the galaxy mass distributions at low radii, since we used extrapolations of the SPARC luminosity distributions at these low distances. As a consequence, the decreasing values for $D\left (R\right )$ at the lowest distances in the top-panel graphs are probably unphysical. Apart from this low-R behavior, the dimension plots for NGC 7814 are very similar to those obtained in papers I-II for general models with spherical symmetry (for example, see the Plummer model in Fig. 3 of paper I and Fig. 4 of paper II). In these previously published figures, we assumed $D =3$ at the origin ($w_{R} =0$), while in this paper we do not force the Newtonian behavior at the galactic center.

Since NGC 7814 is dominated by its spherical bulge, it is not unexpected to obtain plots for $D\left (R\right )$ similar to previously analyzed spherical cases, in which $D \approx 3$ near the center, then $D$ decreases slowly to smaller values at larger distances. For these structures, the
$D \approx 2$ deep-MOND regime is achieved only at very large distances, beyond the last SPARC data point, so this asymptotic behavior is not fully shown in the top panel of Fig. \ref{figure:NGC7841_1}, but we checked it by further extrapolation of the NFDG dimension plots at higher distances than those shown in the figures.

In the bottom panel of the same figure, we produce a detailed fitting to the NGC 7814 rotational velocity data, as given by the related SPARC data-points. We still use here the physical radial distance $R$ in \textrm{kiloparsec}, while rotational circular velocities $v_{circ}$ are measured in $\mbox{km}\ \mbox{s}^{ -1}$.

In particular, in this panel we show the NFDG (MOND) and NFDG (SPARC) curves (blue-solid and green-solid, respectively) which have been extrapolated below $R_{\min } \simeq 0.70\ \ kpc$, due to the convergence issues already mentioned. These curves are also limited by $R_{\max } \simeq 19.53\ kpc$, which is the radial distance of the last SPARC data-point. Again, these limits are shown as vertical thin-gray lines in the figure. In addition, a third NFDG curve (red-dashed) is shown for a fixed value ($D =2$) of the space dimension. This curve can be computed even at very low radial distances, because it is based on the logarithmic potential in the second line of Eq. (\ref{eq2.1}), so it does not suffer from the numerical limitations at low-$R$ of the other NFDG curves.

For this panel we assumed a total mass $M =8.45 \times 10^{40}\thinspace \mbox{kg}$
(obtained by integrating the interpolated total mass distribution),           with
$l_{0} \approx \sqrt{\frac{GM}{a_{0}}} \simeq 2.17 \times 10^{20}\mbox{m}$, and disk scale length $R_{d} =2.54\ \ensuremath{\operatorname*{kpc}} =7.84 \times 10^{19}\ \mbox{m}$  \citep{Lelli:2016zqa} (rescaled length $W_{d} =R_{d}/l_{0} =0.362$). The NFDG circular speeds are computed as $v_{circ} =\sqrt{(g_{obs})_{NFDG}\left (R ,D\left (R\right )\right )R}/10^{3}\left [\mbox{km}\mbox{s}^{ -1}\right ]$, with the dimension functions $D\left (R\right )$ as plotted in the top panel, while the Newtonian speeds are computed as $v_{circ} =\sqrt{(g_{bar})_{SPARC}\left (R\right )R}/10^{3}\left [\mbox{km}\mbox{s}^{ -1}\right ]$, with $(g_{bar})_{SPARC}\left (R\right )$ representing the functions obtained by interpolating the SPARC data for the different components and the total Newtonian one.

The three NFDG curves can be compared with the SPARC data-points (black circles) and related error bars obtained from the published data \citep{Lelli:2016zqa,Lelli:2020pri}, and also with the flat rotation velocity
$V_{f} =218.9 \pm 7.0\left [\mbox{km}\thinspace \mbox{s}^{ -1}\right ]$  \citep{Lelli:2016zqa}, represented by the horizontal gray lines/band. For completeness, we also show the SPARC\  data for the Newtonian cases (disk, gas, bulge, and total Newtonian - gray circles) together with the Newtonian curves (in gray for the different components; black-dashed for the total Newtonian) from the interpolated mass distributions (derived from the original luminosity distributions \citep{Lelli:2020pri}). It is evident from the Newtonian curves of the three components that this galaxy is bulge-dominated, while the disk and gas contributions are less important, even at larger radial distances.

For this bulge-dominated spiral galaxy, the flattening effect of the observed rotation curve is noticeable over most of the radial range. Our NFDG (SPARC) curve (green-solid line) can perfectly model the published data over the applicable range ($R_{\min } ,R_{\max }$) and even at the lower radial values, where the NFDG\ results have been extrapolated. Again, this green-solid curve is obtained by assuming the variable dimension $D\left (R\right )$ as described by the corresponding green-solid curve in the top-panel of Fig. \ref{figure:NGC7841_1}, i.e., assuming that NFG 7841 behaves as a fractal medium whose fractional dimension is described by this function $D\left (R\right )$.

The  NFDG (MOND) curve (blue-solid line) is less effective than the previous one in modeling the SPARC data, but still close to the gray band of the flat rotation velocities over most of the radial range. As already remarked, this curve corresponds to the general RAR, which is an empirical fit to all SPARC\  data-points and, therefore, less accurate in the analysis of an individual galaxy. This shows that, from the NFDG point of view, the RAR\ is only an approximation, while the correct way to model the transition from the Newtonian to the MOND\ regime is to assume a variable dimension $D\left (R\right )$ which is a distinctive characteristic of each galaxy being studied. 

In the bottom panel, we also plot the third NFDG curve (red-dashed) for a fixed value ($D =2$) because this can be easily computed even at low values of $R$ and represents the deep-MOND general limit. This $D =2$ curve is not effective at all in describing NGC 7814 (although remarkably flat over most of the radial range). This is due to the dominant spherical symmetry of this galaxy, which is not consistent with a dimension $D \approx 2$, except perhaps at distances beyond the last SPARC data point.

We interpret the above results as a possible indication that, for bulge-dominated galaxies like NGC 7814, the fractal dimension function $D\left (R\right )$ essentially follows the general behavior of the spherically-symmetric models as studied in paper I: near the galactic center Newtonian regime ($D \simeq 3$) is present, with the dimension $D$ slowly decreasing at larger densities toward the deep-MOND regime $D \simeq 2$ asymptotically. However, over the observed region the dimension remains in the range $2.4 \lesssim D \lesssim 3$, due to the slow decrease of $D$ typical of these spherical structures.  The disk and gas components, which are characterized by dimension values $D \lesssim 2$ (see the next two sub-sections), are not very prominent for NGC 7814 and, therefore, do not play a significant role in this case.

\begin{figure}
	\includegraphics[width=\columnwidth]{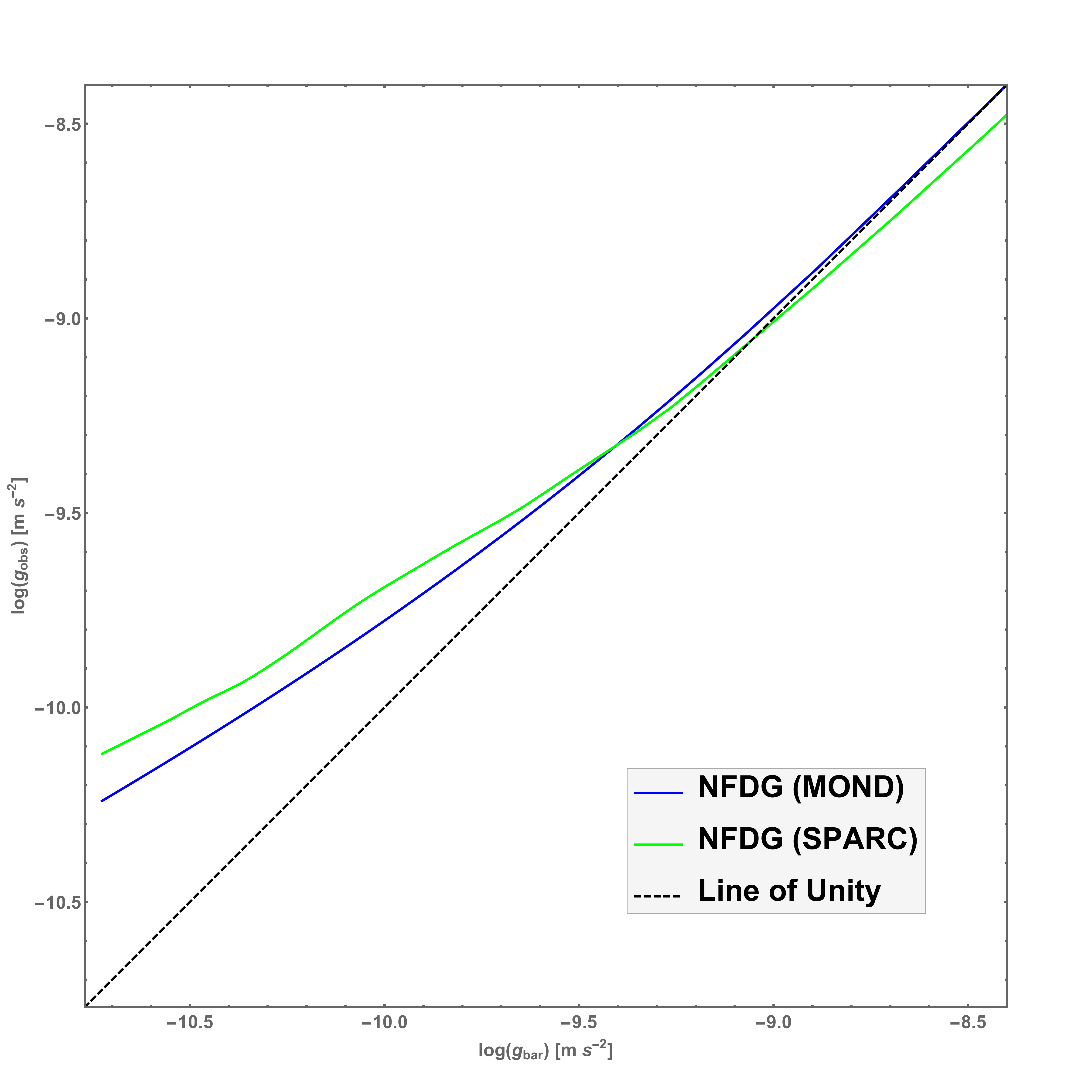}
	\caption{NFDG log-log plots for NGC 7814.
		The NFDG (MOND) curve (blue-solid) represents the general MOND-RAR relation, while the NFDG (SPARC) curve (green-solid) represents the particular case of NGC 7814, since it is based directly on the SPARC data for this galaxy. Also shown: Newtonian behavior-Line of Unity (black-dashed diagonal line).}
	\label{figure:NGC7841_2}
\end{figure}

To conclude the analysis of NGC 7814, Fig. \ref{figure:NGC7841_2} presents the related
$\log \left (g_{obs}\right )$
vs.
$\log \left (g_{bar}\right )$
plots, similar to those used in the literature (see \citet{McGaugh:2016leg}, \citet{Lelli:2017vgz}, and also papers I-II) to illustrate the validity of the general MOND-RAR relation in Eq.
(\ref{eq2.3}). Compared to the \textit{Line of Unity} (black-dashed),
representing the purely Newtonian case, in this figure we show the log-log plots obtained with our NFDG\ models, using
$g_{obs}\left (w_{R}\right )$
based on the dimension functions $D\left (R\right )$
shown in the top panel of Fig. \ref{figure:NGC7841_1} (blue-solid based on MOND-RAR, green-solid based on SPARC data) and with
$g_{bar}\left (w_{R}\right )$
based on Newtonian SPARC data.

The former log-log plot (blue-solid) is equivalent to the general RAR, representing the most general empirical relation for all the galaxies of the SPARC database. The latter log-log plot (green-solid) represents the actual
$\log \left (g_{obs}\right )$
vs.
$\log \left (g_{bar}\right )$
curve for this particular galaxy. The upper-right portion of the figure corresponds to the low-R region, but we have excluded from this plot the extrapolated  region ($0 <R <R_{\min }$), because it is not fully reliable.

From the central part of the figure, toward the lower-left corner (high-R region), the two log-log curves show similar behavior, once they become separated from the Newtonian Line of Unity. However, there are differences between the general NFDG (MOND) plot and the particular NFDG (SPARC) one. Once again, this shows that the RAR is a good general empirical approximation, but each galactic case can be explained in NFDG in terms of its particular dimension function $D\left (R\right )$.


Comparing these first two figures with the similar ones in papers I-II, we note that this time we did not plot the corresponding curves based directly on the
MOND (RAR) or on the\ SPARC data, without any use of NFDG predictions (which appeared as dotted curves in all the similar plots of papers I-II). These dotted curves were used in papers I-II to double-check our NFDG results and, in general, they were completely consistent with the solid curves of the NFDG predictions. We checked this perfect correspondence also for the situations analyzed in this paper, but we omitted these additional curves as they were not providing any additional information to our analysis.

\subsection{NGC 6503}
\label{subsect:NGC6503}

As our second galaxy, we revisit the case of NGC 6503 which has been already studied in a preliminary way in our paper II \citep{Varieschi:2020dnd}. Here we expand our previous analysis of this galaxy and compare the NFDG results with the other cases studied in this paper. NGC 6503 is a field dwarf spiral galaxy approximately 17 million light-years away in the constellation of Draco. This spiral galaxy shows bright blue regions typically related to star formation, bright red regions of gas along its spiral arms, and dark-brown dust areas in the galaxy's center and arms, as seen in images taken by the NASA/ESA Hubble Space Telescope \citep{NGC6503:2020}.

This is a disk-dominated galaxy, in which only two of the three components are present - disk and gas - with the axially-symmetric stellar disk component being the most prominent. Following the general discussion of NFDG methods outlined in Sect. \ref{sect:REVIEW}, we converted the SPARC surface luminosity data for the two components into an equivalent mass distribution $\widetilde{\Sigma }\left (w_{R}^{ \prime }\right )$ (disk plus gas) and entered this function into our general NFDG thick-disk potential in Eqs. (\ref{eq2.12})-(\ref{eq2.13}), from which we then obtained $\mathbf{}g_{obs}\left (w_{R}\right )$ by differentiation. Since there is no spherical bulge for this galaxy, we did not include any contribution from Eqs. (\ref{eq2.14})-(\ref{eq2.15}) in this case.

\begin{figure}
	\includegraphics[width=\columnwidth]{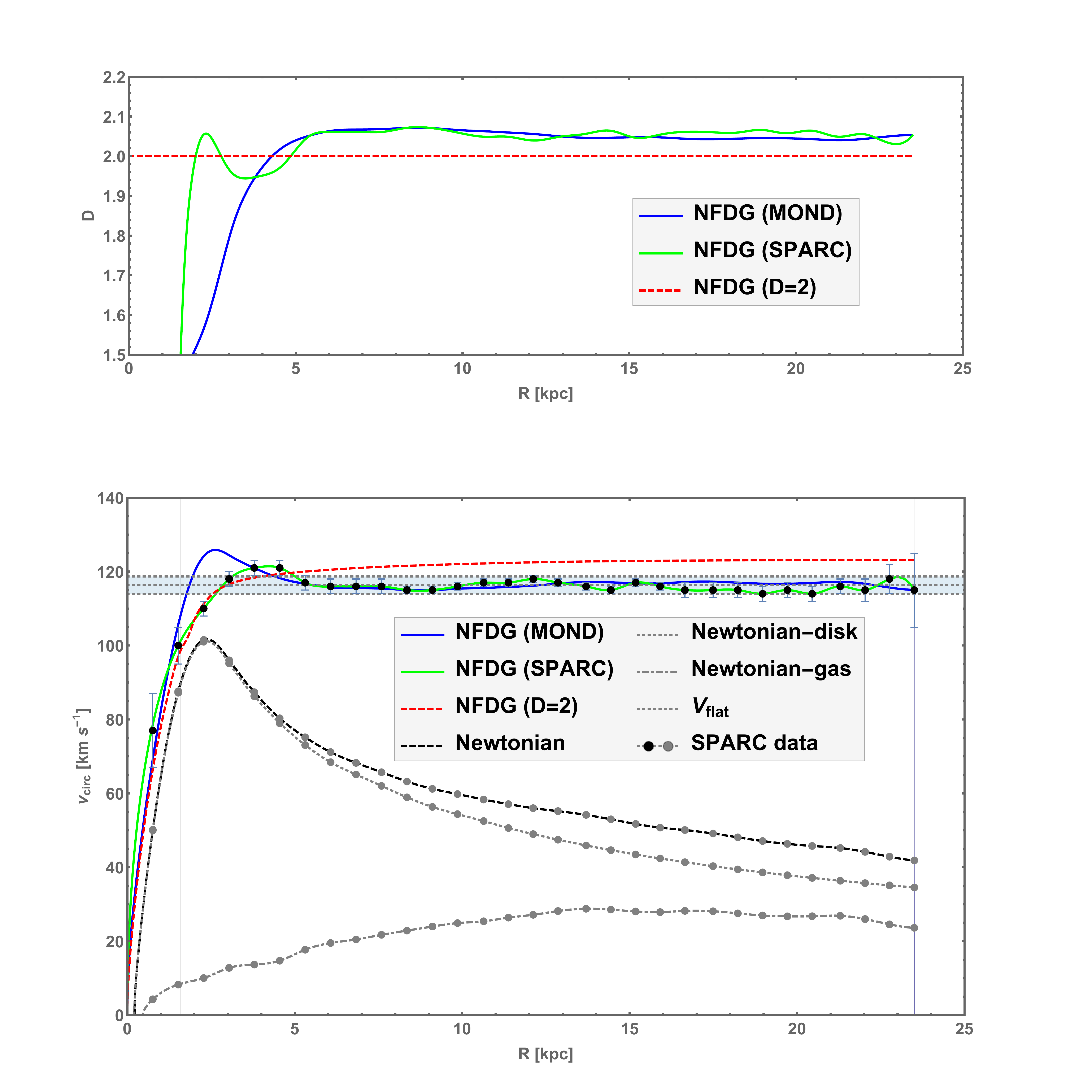}
	\caption{NFDG results for NGC 6503.
		Top panel: NFDG variable dimension $D\left (R\right )$ for MOND (RAR) interpolation function, or based directly on SPARC data. Bottom panel: NFDG rotation curves (circular velocity vs. radial distance) compared to the original SPARC data (black circles). Also shown: Newtonian rotation curves (total and different components - gray lines) and corresponding SPARC data (gray circles).}
	\label{figure:NGC6503_1}
\end{figure}

Figure \ref{figure:NGC6503_1} summarizes the main NFDG results for NGC 6503, in the same way of the similar Fig. \ref{figure:NGC7841_1} in the previous subsection. The top panel shows the variable dimension function $D\left (R\right )$ computed using our two main options (based on the RAR or on SPARC data) over the range for the radial distance $R$ consistent with available observational data. The bottom panel shows the NFDG circular speeds for our two main options (blue-solid and green-solid) as well as the curve for fixed $D =2$ (red-dashed), compared with SPARC data and with the NGC 6503 flat rotation velocity
$V_{f} =116.3 \pm 2.4\left [\mbox{km}\thinspace \mbox{s}^{ -1}\right ]$  \citep{Lelli:2016zqa} represented by the horizontal gray lines and gray band in the figure. Newtonian data and curves are also included, for completeness.

For this galaxy, we assumed a total mass $M =1.72 \times 10^{40}\thinspace \mbox{kg}$ with
$l_{0} \approx \sqrt{\frac{GM}{a_{0}}} \simeq 9.79 \times 10^{19}\mbox{m}$, and disk scale length $R_{d} =2.16\ \ensuremath{\operatorname*{kpc}} =6.67 \times 10^{19}\ \mbox{m}$ (rescaled length $W_{d} =R_{d}/l_{0} =0.681$) \citep{Lelli:2016zqa}. As in the previous case, results are extrapolated below $R_{\min } \simeq 1.59\ \ kpc$, due to the convergence issues already mentioned, although we improved on these results at low-R in this work, compared to those presented in our paper II. These curves are also limited by $R_{\max } \simeq 23.5\ kpc$, which is the radial distance of the last SPARC data-point. The limits are shown as usual by the vertical thin-gray lines in the two panels of Fig. \ref{figure:NGC6503_1}.


As already remarked in paper II, the top panel shows $D\left (R\right ) \simeq 2$ for our main NFDG results over most of the radial range, with the exception of the values at low-R. We interpret these results as a strong indication that, for disk-dominated galaxies such as NGC 6503, the fractal dimension over the whole radial range is approximately $D \simeq 2$, with just small variations from this almost constant value. This also follows from our heuristic arguments, presented in this paper as well as in papers I-II, that the MOND\ behavior is essentially related to an effective Newtonian gravity in a space of reduced dimension $D \simeq 2$. Thick disk galaxies, with a dominant stellar disk component, are probably more likely to show this behavior, as opposed to bulge and gas-dominated galaxies characterized by different dimension functions.

The bottom panel in Fig. \ref{figure:NGC6503_1} confirms our analysis above, by showing a perfect agreement of our main NFGD (SPARC) fit (green-solid) with the SPARC data, even in the low-R extrapolated range. The more general NFDG (MOND) curve (blue-solid) also fits most data reasonably well, showing that the RAR is more effective for this type of galaxies. The fixed $D =2$ NFDG curve, which does not suffer from any extrapolation at low-R, is able to fit well the low distance data, while is less effective at higher distances, but still remarkably flat. In summary, NFDG applied to NGC 6503 shows a perfect example of MOND\ behavior explained by an almost constant $D \simeq 2$ fractal dimension.

\begin{figure}
	\includegraphics[width=\columnwidth]{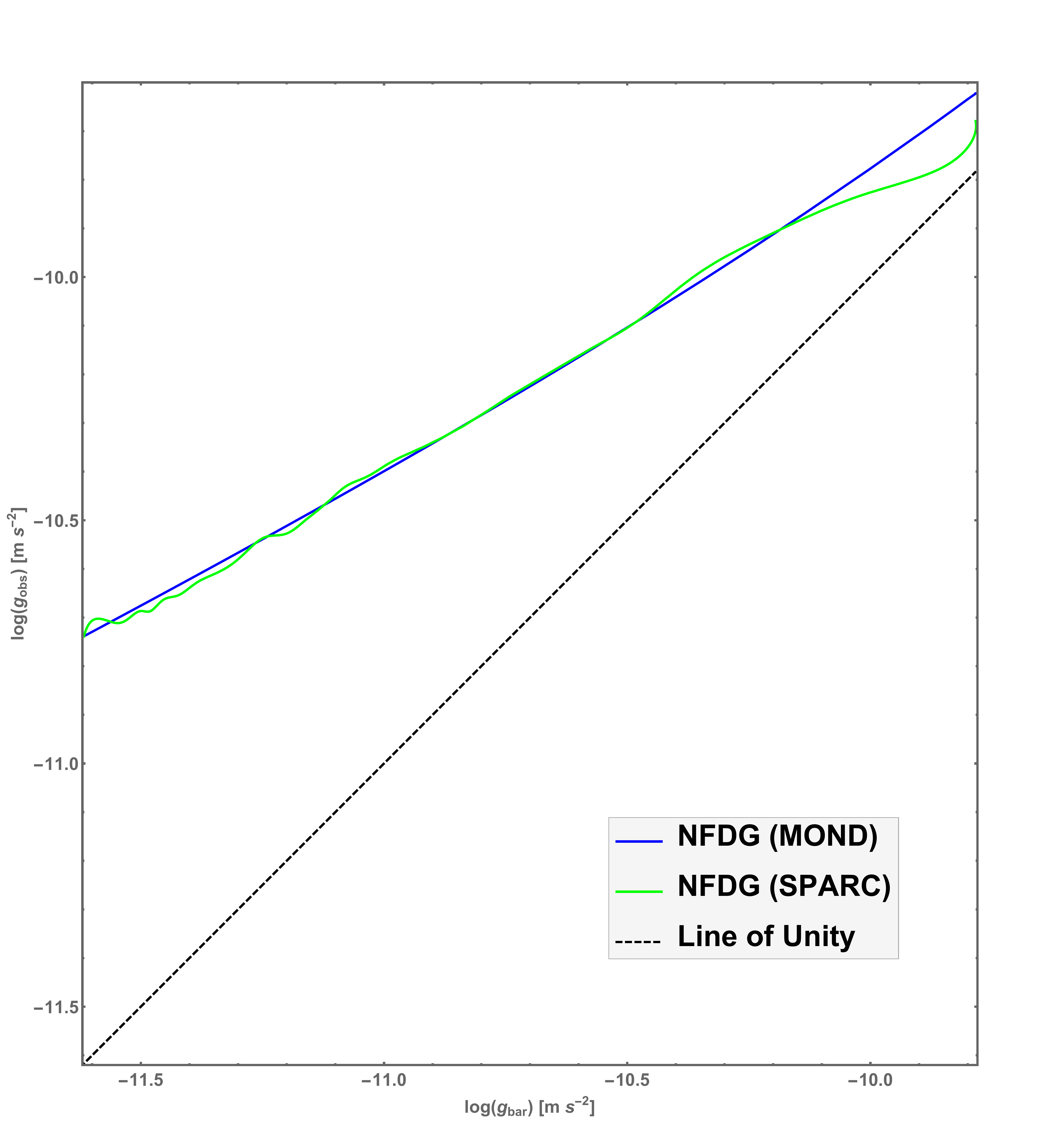}
	\caption{NFDG log-log plots for NGC 6503.
		The NFDG (MOND) curve (blue-solid) represents the general MOND-RAR relation, while the NFDG (SPARC) curve (green-solid) represents the particular case of NGC 6503, since it is based directly on the SPARC data for this galaxy. Also shown: Newtonian behavior-Line of Unity (black-dashed diagonal line).}
	\label{figure:NGC6503_2}
\end{figure}

In figure \ref{figure:NGC6503_2}, we also show the
$\log \left (g_{obs}\right )$
vs.
$\log \left (g_{bar}\right )$
plots for NGC 6503, similar to those shown in Fig. \ref{figure:NGC7841_2} for NGC 7814. As usual the blue-solid curve represents the general MOND-RAR case, while the green-solid curve is peculiar to the individual galaxy being studied. Apart from the results shown in the top-right corner of this figure (representing the low-R region, still excluding the extrapolated part for $0 <R <R_{\min }$), there is strong agreement between the general MOND-RAR curve and the particular NFDG-SPARC one. This shows again that disk-dominated galaxies are probably closer to the general behavior described by the RAR\ relation, as compared with bulge and gas-dominated galaxies.


\subsection{NGC 3741}
\label{subsect:NGC3741}

In our third and final case we consider NGC 3741, an irregular galaxy discovered by J. Herschel in 1828. It is a gas-dominated spiral galaxy approximately 11 million light-years away in the constellation Ursa Major \citep{NGC3741:2020}. Only two of the three components are present - disk and gas - with the axially-symmetric gas component being the most prominent. Following again the general discussion of NFDG methods in Sect. \ref{sect:REVIEW}, we converted the SPARC surface luminosity data for the two components into an equivalent mass distribution $\widetilde{\Sigma }\left (w_{R}^{ \prime }\right )$ (disk plus gas) and entered this function into our general NFDG thick-disk potential in Eqs. (\ref{eq2.12})-(\ref{eq2.13}), from which we then obtained $\mathbf{}g_{obs}\left (w_{R}\right )$ by differentiation.

\begin{figure}
	\includegraphics[width=\columnwidth]{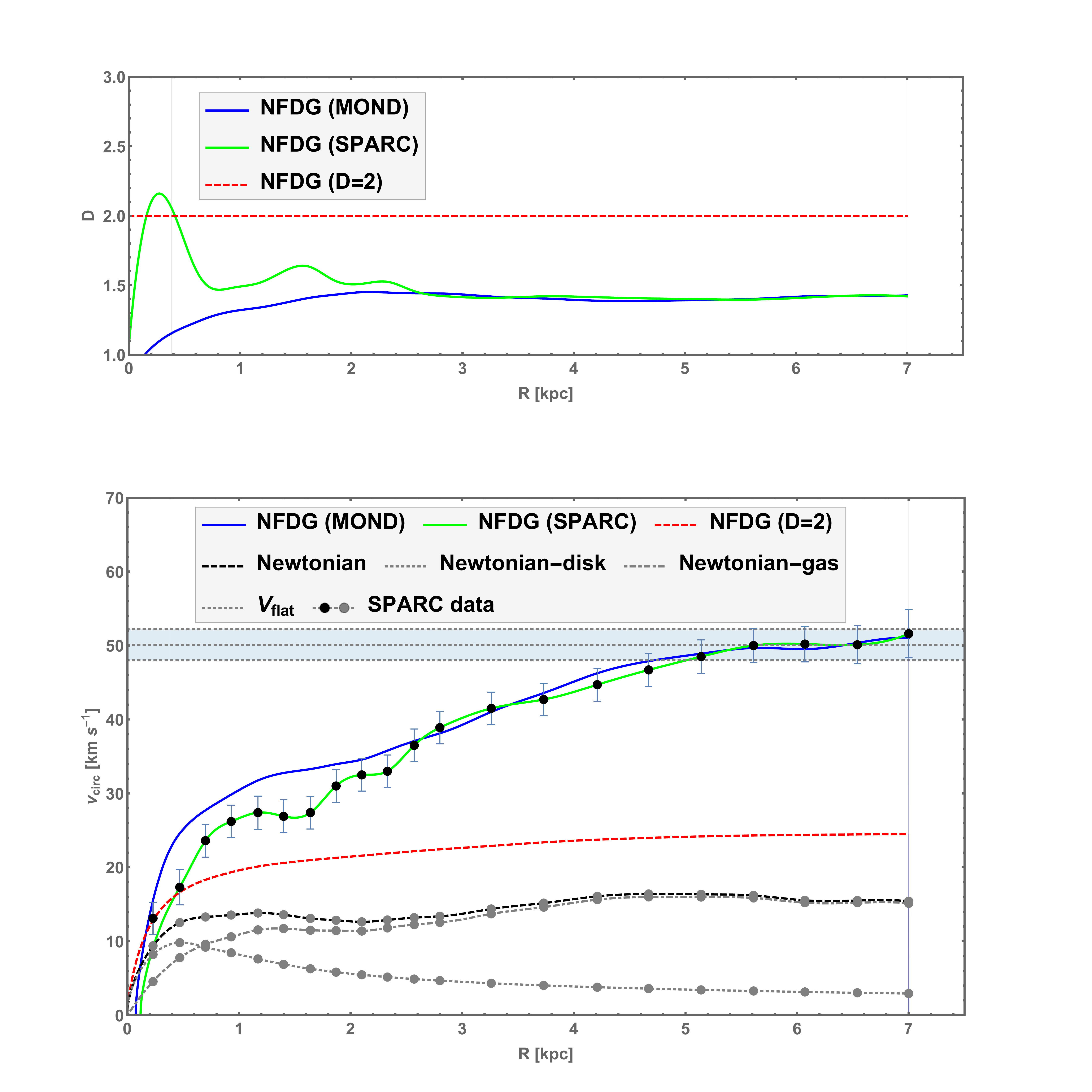}
	\caption{NFDG results for NGC 3741.
		Top panel: NFDG variable dimension $D\left (R\right )$ for MOND (RAR) interpolation function, or based directly on SPARC data. Bottom panel: NFDG rotation curves (circular velocity vs. radial distance) compared to the original SPARC data (black circles). Also shown: Newtonian rotation curves (total and different components - gray lines) and corresponding SPARC data (gray circles).}
	\label{figure:NGC3741_1}
\end{figure}

Figure \ref{figure:NGC3741_1} summarizes the main NFDG results for NGC 3741, in the same way of the similar figures in the previous subsections. The top panel shows the variable dimension function $D\left (R\right )$ computed using our two main options, while the bottom panel shows the NFDG circular speeds for the same options (blue-solid and green-solid) as well as the curve for fixed $D =2$ (red-dashed), compared with SPARC data and with the NGC 3741 flat rotation velocity
$V_{f} =50.1 \pm 2.1\left [\mbox{km}\thinspace \mbox{s}^{ -1}\right ]$  \citep{Lelli:2016zqa} represented by the horizontal gray lines and gray band in the figure. Newtonian data and curves are also included, for completeness.

For this galaxy, we assumed a total mass $M =5.06 \times 10^{38}\thinspace \mbox{kg}$ with
$l_{0} \approx \sqrt{\frac{GM}{a_{0}}} \simeq 1.68 \times 10^{19}\mbox{m}$, and disk scale length $R_{d} =0.20\ \ensuremath{\operatorname*{kpc}} =6.17 \times 10^{18}\ \mbox{m}$ (rescaled length $W_{d} =R_{d}/l_{0} =0.368$) \citep{Lelli:2016zqa}. As in the previous cases, our results are extrapolated below $R_{\min } \simeq 0.38\ \ kpc$ and also limited by $R_{\max } \simeq 7.00\ kpc$, which is the radial distance of the last SPARC data-point. The limits are shown as usual by the vertical thin-gray lines in the two panels of Fig. \ref{figure:NGC3741_1}.


The top panel shows $D\left (R\right ) \simeq 1.4 -1.6$ for our main NFDG results over most of the radial range, with the exception of the values at low-R. These results suggest that, for gas-dominated galaxies such as NGC 3741, the fractal dimension over most of the radial range is $1 <D \lesssim 2$, which is notably different from the first two cases where $2 \lesssim D \leq 3$. Thick disk galaxies with a dominant gas component might be characterized by a lower fractal dimension and might attain the deep-MOND ($D \approx 2$) regime only at very large radial distances, beyond the observed range.

The bottom panel in Fig. \ref{figure:NGC3741_1} supports our analysis, by showing a perfect agreement of our main NFGD (SPARC) fit (green-solid) with the SPARC data, except in the very low-R extrapolated range, where the NFDG $D =2$ curve fits the first experimental point better. The more general NFDG (MOND) curve (blue-solid) also fits most data in the mid-high range well, but is much less effective at low-R, showing that the RAR is just an approximation for this type of galaxies. 

The fixed $D =2$ NFDG curve, which does not suffer from any extrapolation at low-R, is unable in this case to fit the data (except the first two points), since the flattening of the circular velocity data only happens at high-R values, for the last four data points. In summary, NFDG applied to NGC 3741 shows a more complex behavior, compared to the previous two cases, probably due to the fractal effects of the gas component.

\begin{figure}
	\includegraphics[width=140mm]{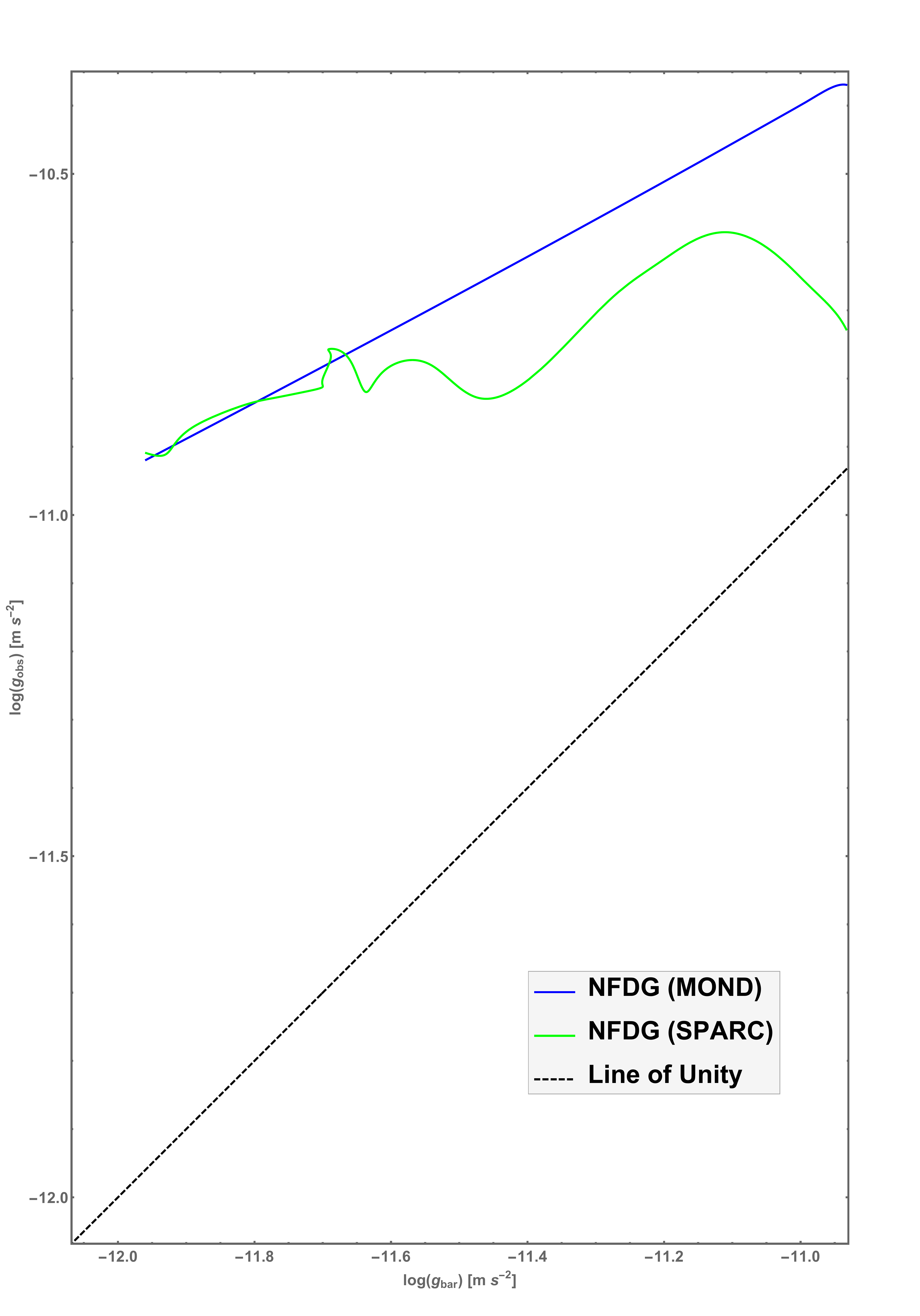}
	\caption{NFDG log-log plots for NGC 3741.
		The NFDG (MOND) curve (blue-solid) represents the general MOND-RAR relation, while the NFDG (SPARC) curve (green-solid) represents the particular case of NGC 3741, since it is based directly on the SPARC data for this galaxy. Also shown: Newtonian behavior-Line of Unity (black-dashed diagonal line).}
	\label{figure:NGC3741_2}
\end{figure}

In figure \ref{figure:NGC3741_2}, we show the
$\log \left (g_{obs}\right )$
vs.
$\log \left (g_{bar}\right )$
plots for NGC 3741, similar to those for the previous galaxies. Again, the blue-solid curve represents the general MOND-RAR case, while the green-solid curve is peculiar to NGC 3741. The results shown in the top-right corner of this figure represent the low-R region, and the MOND and SPARC plots do not match in this region. The two NFDG curves agree well only in the left part of the figure (high-R region), generally showing that the RAR is not a good approximation for this gas-dominated galaxy.


\subsection{Discussion}
\label{subsect:discussion}

Following the three cases analyzed in the previous sub-sections, we can attempt to generalize our NFDG results to all cases of rotationally supported galaxies. While the RAR is an effective empirical law which works for most of the galaxies in the SPARC database and others, we argued that it is only an approximation of the fractal behavior of these galaxies. This behavior is characterized by the variable fractional dimension function $D\left (R\right )$, as observed in the galactic plane ($z =0$) and thus depending only on the radial distance, for symmetry reasons.

This dimension function $D\left (R\right )$ is an individual feature of each galaxy that can be derived directly from the observed circular velocities and luminosity distributions, properly converted into mass distributions for each galaxy being considered. In general, the fractional dimension range is $1 <D\left (R\right ) \leq 3$, as seen in the three cases studied in this work. Once the full dimension function has been determined, the NFDG formulas for the gravitational potentials and fields are able to produce accurate fits to the observed circular speed data, without the need of any dark matter or other supplemental hypotheses.

Although the dimension function $D\left (R\right )$ is peculiar to each galaxy analyzed, it is likely that the interplay of the three main components - bulge, disk, and gas - will play a significant role in determining this dimension function. We have argued that if one of these three components is dominant, as in the galaxies studied in this work, it might determine the overall shape of the $D\left (R\right )$ function. 

Bulge-dominated galaxies, such as NGC 7814, will probably be characterized by dimension functions which are typical of the spherically-symmetric structures analyzed in paper I: $D$ slowly decreasing from the Newtonian $D \approx 3$ in the central region toward $D \approx 2.4 -2.6$ in the outer regions, and then asymptotically $D \approx 2$ at very large distances. The related rotation velocity curves would become flat at larger radial distances, where the different components become comparable (see Fig. \ref{figure:NGC7841_1}). The RAR is an effective approximation in this case, but not perfect at larger distances (see Fig. \ref{figure:NGC7841_2}).

Disk-dominated galaxies, such as NGC 6503, are more likely to exhibit deep-MOND behavior, and thus $D \approx 2$, over most of the radial range. As a consequence, they also show a remarkably flat circular speed curve over the bulk of the radial range (see Fig. \ref{figure:NGC6503_1}) and they follow more closely the general RAR (see Fig. \ref{figure:NGC6503_2}). They show the most typical MOND behavior, related to the NFDG $D \approx 2$ value for the fractal dimension. 

Gas-dominated galaxies, such as NGC 3741, show a more complicated NFDG behavior. The dominant gas component possibly drives the dimension to lower values $1 <D \lesssim 2$ over nearly all the radial range. As a result, the circular velocity plot is not flat over most distances and an asymptotic speed is only achieved at the largest distances (see Fig. \ref{figure:NGC3741_1}). In this case, the RAR is less effective in modeling the transition between Newtonian and MOND regimes (see Fig. \ref{figure:NGC3741_2}). Future work on other galaxies from the SPARC database will be needed to confirm if the observed behaviors for the three cases analyzed above are indeed a general feature of NFDG.

In closing this section, we want to comment on the theory by A. Giusti and collaborators \citep{Giusti:2020rul,Giusti:2020kcv} which shares many similarities with NFDG, although it is based on different ``fractional'' assumptions. We already discussed in paper I (see Appendix A in \citet{Varieschi:2020ioh}) how Giusti introduced a fractional version of Newton's theory \citep{Giusti:2020rul} based on a fully fractional Poisson equation, where the ordinary Laplacian $-\Delta $ is replaced by a fractional version $\left ( -\Delta \right )^{s}$ with $s \in [1 ,3/2)$. In this way, setting $s =1$ the Newtonian limit is recovered, while $s =3/2$ corresponds to MOND behavior. Giusti's model has been named \textit{Fractional Newtonian Gravity} (FNG) in a second paper on the subject \citep{Giusti:2020kcv}.

As already remarked in papers I-II and in Sect. \ref{sect:intro} of this work, the main difference between FNG and our NFDG is that the former is a MOND-like (non-local) fractional theory, i.e., fractional operators are used to determine the gravitational potential and field, while the latter employs field equations of integer order (and thus local). The ``fractional flavor'' of NFDG comes instead from the non-integer dimension $D \leq 3$ of the underlying fractal space. Both models are (at the moment) non-relativistic and are also based on a MOND scale length ($\ell  =\frac{2}{\pi }\sqrt{GM/a_{0}}$ for FNG, $l_{0} =\sqrt{GM/a_{0}}$ for NFDG) which ensures dimensional correctness of all the equations.

The gravitational potentials obtained by the two models also present some similarities with regard to their general forms, but they differ in the mathematical details \citep{Varieschi:2020ioh}. Still, they show similarities when plotted for equivalent values of the fractional dimension, so it is possible that these two models might be practically equivalent, when applied to galactic structures. In both FNG papers \citep{Giusti:2020rul,Giusti:2020kcv} Giusti et al., apply their model to the gravitational potential for Kuzmin disks and also consider the possibility of turning FNG into a variable order theory, by assuming $s =s\left (\mathbf{x}/\ell \right )$, i.e., considering the fractional order as a function of the field point. This is equivalent to our NFDG assumption of a variable fractional dimension $D =D\left (\mathbf{w}\right )$, depending on the rescaled field position $\mathbf{w} \equiv \mathbf{r}/l_{0}$. 

It will be interesting to compare the detailed NFDG dimension/rotation curves presented in this paper for the three galaxies that we have analyzed, with similar FNG fits, if they will become available. In particular, after fitting the rotational velocity data for a certain galaxy (e.g., NGC 6503), if the two models will yield the same variable dimension function (i.e., if the FNG $s =s\left (\mathbf{x}/\ell \right )$ function will be found to be equivalent to the NFDG $D =D\left (R\right )$), then the two models could also be considered equivalent and the case for a ``fractional'' explanation of MOND would become stronger.

\section{Conclusions}
\label{sect::conclusion}

In this work, we continued our study of NFDG, a fractional-dimension gravity model which might provide a possible explanation of the MOND theory and the related RAR. As in our papers I-II, we assumed that Newtonian gravity
might act on a metric space of variable dimension
$D \leq 3$, when applied to galactic scales, and used NFDG to model three rotationally supported galaxies: NGC 7814, NGC 6503, and NGC 3741.

For these three cases, we have shown that NFDG can accurately fit the rotation data, as it was done by MOND-RAR models, but these results are explained in terms of dimension functions $D\left (R\right )$ which are peculiar to each of the galaxies being studied. In particular, a bulge-dominated galaxy such as NGC 7814 is characterized by a dimension function typical of spherically symmetric structures ($2 <D \leq 3$), a disk-dominated galaxy such as NGC 6503 is characterized by an almost constant $D \approx 2$ dimension, and a gas-dominated galaxy such as NGC 3741 is characterized by lower values for the space dimension ($1 <D \lesssim 2$). The deep-MOND\ regime, $D \approx 2$, is more typical of disk-dominated galaxies, while it is achieved only asymptotically by bulge and gas-dominated galaxies.

Further work will be needed to check this general interpretation of the results. Detailed fitting of several other galaxies in the SPARC database will need to be performed in upcoming work to show how the three different components (bulge, disk, and gas) play different roles in determining the dimension function for each case. Additionally, other structures for which MOND is less effective, such as globular clusters or similar, will need to be studied to see if their dynamical behavior can still be explained by NFDG without any use of DM. 

A relativistic version of NFDG, which can be tentatively called Relativistic Fractional-Dimension Gravity (RFDG) also needs to be introduced, by expanding General Relativity to metric spaces with fractional dimension. This relativistic version might also lead to a possible explanation for the origin of the variable dimension of each galactic structure. We will leave these and other topics to future work on the subject.

\section*{Acknowledgements}

This work was supported by a Faculty Sabbatical Leave granted by Loyola Marymount University, Los Angeles. The author also wishes to acknowledge Dr. Federico Lelli for sharing SPARC galactic data files and other useful information, Dr. Gianluca Calcagni for helpful advice regarding multifractional theories, and the anonymous reviewer for useful comments and suggestions.

\section*{Data Availability}

The data underlying this article will be shared on reasonable request to the corresponding author.



\bibliographystyle{mnras}
\bibliography{mainNotes} 




\appendix

\section{NFDG and variable dimension D}
\label{sect::appendix}

In this section, we discuss in a more detailed way possible mathematical techniques for achieving a variable dimension $D$ in the context of the NFDG model.

It should be noted again that the NFDG main equation (\ref{eq2.1}), and the other fundamental equations introduced in papers I-II, were derived for a fixed value of the dimension $D$, while in general we assume a variable dimension $D(w_{R})$. In papers I-II, we justified this transition from constant to variable dimension by assuming a slow change over galactic distances of this varying dimension, so that the fundamental NFDG equations were still approximately valid when replacing $D$ with $D(w_{R})$. However, this transition from constant to varying dimension should be introduced in a more rigorous way. In the following, we outline two different ways, but others might be also feasible.

In fractional calculus, integration and differentiation of variable order has been studied in several papers \citep{doi:10.1080/10652469308819027,Samko:1995a,Samko:1995b,Samko:2005,Samko:2013}, where the order of integration/differentiation can be a function of the space point. In particular, (left-sided) Riemann-Liouville fractional integrals were considered with variable order $\alpha(x)>0$ \citep{doi:10.1080/10652469308819027,Samko:1995a,Samko:1995b}:
\begin{equation}I_{a,x}^{\alpha(\cdot) }f(x) =\frac{1}{\Gamma [\alpha(x) ]}\int _{a}^{x}(x -t)^{\alpha(x)  -1}f(t)dt , \label{eq5.1}
\end{equation}
where $a\geq-\infty$, $x>a$, and the variable order of integration may vary from point to point. This theory was then extended even to multi-dimensional cases and fractional spherical integrals \citep{Samko:2013}, showing that this approach is mathematically sound.

It might be possible to extend this theory to the right-sided version of the integral in the previous equation, $I_{x,a}^{\alpha(\cdot) }f(x) =\frac{1}{\Gamma [\alpha(x)] }\int _{x}^{a}(t -x)^{\alpha(x)  -1}f(t)dt$, and eventually introduce also a variable-order Weyl's fractional integral of the form: $W^{ -D(\cdot)}f(0) =\frac{1}{\Gamma [D(\cdot)]}\int _{0}^{\infty }t^{D(\cdot) -1}f(t)dt$, although the current mathematical theory does not include these integrals. We recall that similar integrals were introduced, for a constant dimension $D$, in Eqs. (8)-(9) of paper I, here combined together as:
\begin{equation}\int _{W}fd\mu _{H} =\frac{2\pi ^{D/2}}{\Gamma (D/2)}\int _{0}^{\infty }f(r)r^{D -1}dr=\frac{2\pi ^{D/2}\Gamma (D)}{\Gamma (D/2)}W^{ -D}f(0). \label{eq5.2}
\end{equation}

In paper I, we based the NFDG model on the above integral, for a space of constant dimension $D$. By analogy with Eq. (\ref{eq5.1}), if we could extend also Eq. (\ref{eq5.2}) to a variable dimension $D(\cdot)$, with a possible dependence on the field point, we would be able to justify the transition from constant to variable dimension in NFDG. Again, the mathematical theory outlined in \citet{Samko:2013} and references therein is far from being complete, thus requiring future additional studies on the matter.

Another way to extend NFDG into a variable-dimension model is to follow existing studies in the literature on multiscale spacetime/multi-fractional theories \citep{Calcagni:2011sz,Calcagni:2016azd,Calcagni:2018dhp,Carlip:2019onx,Calcagni:2016xtk}. 
In quantum gravity theories, the dimension of space or spacetime may change with the scale of observation (dimensional flow or multiscaling). In \citet{Calcagni:2016xtk} two general theorems were introduced for this dimensional flow. Under very general assumptions, the first flow-equation theorem describes the flow of the spacetime (Hausdorff or spectral) dimension $d(\ell)$ as a function of the scale parameter $\ell$:
\begin{equation}d(\ell)\simeq D+b\left(\frac{\ell_{*}}{\ell}\right)^{c}+(log\ oscillations), \label{eq5.3}
\end{equation}
where $\ell_{*}$ is a reference scale separating the infrared (IR) properties of the flow from the ultraviolet (UV) properties, and $b$, $c$ are some parameters to be determined.

In the second flow-equation theorem, the overall measure of the spacetime is assumed to be factorizable, $d=\sum_{\mu} d^{(\mu)}$, where for each direction $d^{(\mu)}$ is given by Eq. (\ref{eq5.3}) with $D=1$, $b\rightarrow b_{\mu}$, and $c\rightarrow c_{\mu}$. In the case of $d=d_H$ (Hausdorff dimension), the most general real measure for each direction of the spacetime is given by the binomial measure \citep{Calcagni:2016azd}:
\begin{equation}q_{\alpha}(x)=x+\frac{\ell_{*}}{\alpha}sgn(x)\left|\frac{x}{\ell_{*}}\right|^{\alpha}F_{\omega}(x), \label{eq5.4}
\end{equation}
with a modulation factor $F_{\omega}(x)=1+(log\ oscillations)$. 

Neglecting these oscillations ($F_{\omega}(x)\simeq1$) and differentiating the measure in Eq. (\ref{eq5.4}), we obtain $dq_{\alpha}(x)=\left(1+\left|\frac{x}{\ell_{*}}\right|^{\alpha-1}\right)dx$, a one-dimensional measure which is practically equivalent to the one used in NFDG for each space direction $i$: $d\mu _{i}(x_{i}) =\frac{\pi ^{\alpha _{i}/2}}{\Gamma (\alpha _{i}/2)}\left \vert \frac{x_{i}}{l_{0}}\right \vert ^{\alpha _{i} -1}dx_{i} ,\ i =1 ,2 ,3$, with $l_0$ representing the NFDG scale length and the total space dimension $D=\alpha_1+\alpha_2+\alpha_3$. The only difference between these two measures would be the overall numerical prefactor in the NFDG measure and the numerical addend in the binomial measure, which is needed to reduce multiscale geometries to standard ones at scales much greater than $\ell_*$  \citep{Calcagni:2021pri}.

The NFDG measure yields the result in Eq. (\ref{eq5.2}), when used to integrate a spherically-symmetric function $f(r)$ over a D-dimensional space. Therefore, we might identify the reference scale lengths of the two models as $\ell_* \equiv l_0$, the scale parameter with the radial distance, $\ell \equiv r$, and use the dimension flow in Eq. (\ref{eq5.3}), where the Hausdorff dimension $d(\ell)$ can now be considered the NFDG varying dimension $D(w_{r})$ for spherically-symmetric cases ($w_r=r/l_0$). Assuming that the two angular coordinates ($\theta$ and $\varphi$) simply add a constant unit dimension to the right-hand side of Eq. (\ref{eq5.3}), we then have: 
\begin{equation}D(w_{r})\simeq 3+b\left(\frac{l_{0}}{r}\right)^{c}+(log\ oscillations) = 3+b\left(w_r\right)^{-c}+(log\ oscillations). \label{eq5.5}
\end{equation}

Neglecting the log oscillations and with parameters $b<0$, $c<0$, the previous equation might represent a varying dimension with $D \simeq 3$ near the origin (Newtonian regime) and then decreasing at larger radial distances (MOND regime), a behavior similar to the one discussed in papers I-II for spherically-symmetric cases. However, the values of the parameters $b$ and $c$ in Eq. (\ref{eq5.5}), as well as possible log oscillations, should be determined from the dynamics of the system being considered \citep{Calcagni:2016azd}. 

As mentioned in Sect. \ref{sect::conclusion} above, at the moment NFDG is a non-relativistic model, with a planned extension into a relativistic version (RFDG). Only when this relativistic model will be established, we might be able to fully connect NFDG/RFDG with the multiscale/multifractional theories that were briefly summarized in this section. The possibility of embedding NFDG, as well as Giusti's model \citep{Giusti:2020rul,Giusti:2020kcv}, into more general fractional and multi-fractional theories has also been recently discussed by \citet{Calcagni:2021new}.


\bsp	
\label{lastpage}
\end{document}